\documentclass[pdflatex,sn-mathphys-num]{sn-jnl}
\usepackage{amsmath}
\usepackage{graphicx}
\usepackage[normalem]{ulem}
\usepackage{svg}
\usepackage{multirow}
\usepackage{pifont}
\usepackage{pax}  
\usepackage{rotating}
\usepackage{tabularray}
\SetTblrTemplate{head}{empty}
\DefTblrTemplate{foot}{default}{%
  \UseTblrTemplate{caption}{default}%
}
\usepackage{anyfontsize}

\usepackage{graphicx}%
\usepackage{multirow}%
\usepackage{amsmath,amssymb,amsfonts}%
\usepackage{amsthm}%
\usepackage{mathrsfs}%
\usepackage[title]{appendix}%
\usepackage{xcolor}%
\usepackage{textcomp}%
\usepackage{manyfoot}%
\usepackage{booktabs}%
\usepackage{algorithm}%
\usepackage{algorithmicx}%
\usepackage{algpseudocode}%
\usepackage{listings}%

\usepackage{xurl}

\usepackage{fancyvrb}
\usepackage{marginnote}
\newcommand{\hlight}[2]{#2}
\newcommand{\cmark}{\ding{51}}%
\newcommand\ccmark{\cmark\kern-0.4em\cmark}

\begin{document}

\title[Detecting and Mitigating DDoS Attacks with AI: A Survey]{Detecting and Mitigating DDoS Attacks with AI: A Survey}

\author[1]{\fnm{Alexandru} \sur{Apostu}}
\email{alexandru.m.apostu@gmail.com}
\equalcont{Authors contributed equally to this research.}
\author[1]{\fnm{Silviu} \sur{Gheorghe}}
\email{silviu-florin.gheorghe@unibuc.ro}
\equalcont{Authors contributed equally to this research.}
\author[1]{\fnm{Andrei} \sur{Hîji}}
\email{andrei-iulian.hiji@unibuc.ro}
\equalcont{Authors contributed equally to this research.}
\author[1]{\fnm{Nicolae} \sur{Cleju}}
\email{ncleju@etti.tuiasi.ro}
\author[1]{\fnm{Andrei} \sur{Pătrașcu}}
\email{andrei.patrascu@fmi.unibuc.ro}
\author[1]{\fnm{Cristian} \sur{Rusu}}
\email{cristian.rusu@fmi.unibuc.ro}
\author[1]{\fnm{Radu Tudor} \sur{Ionescu}}
\email{raducu.ionescu@gmail.com}
\author*[1]{\fnm{Paul} \sur{Irofti}}
\email{paul@irofti.net}

\affil[1]{\orgdiv{Department of Computer Science}, 
    \orgname{University of Bucharest},\\
    \orgaddress{\street{Panduri 90},
    \city{Bucharest},
    \postcode{050663},
    \country{Romania}}
}


\abstract{Distributed Denial of Service attacks represent an active cybersecurity research problem.  Recent research shifted from static rule-based defenses towards AI-based detection and mitigation.  This comprehensive survey covers several key topics.  Preeminently, state-of-the-art AI detection methods are discussed.  An in-depth taxonomy based on manual expert hierarchies and an AI-generated dendrogram are provided, thus settling DDoS categorization ambiguities.  An important discussion on available datasets follows, covering data format options and their role in training AI detection methods together with adversarial training and examples augmentation.  Beyond detection, AI based mitigation techniques are surveyed as well. Finally, multiple open research directions are proposed.}


\maketitle


\section{Introduction}

DoS (Denial of Service) and DDoS (Distributed Denial of Service) represent a major cybersecurity threat. DDoS attacks increased by 55\% between January 2020 and March 2021 \cite{Shroff-Wiley-2022}, a phenomenon that was accentuated starting from the second half of 2021 and continued until the first half of 2022, when the total number of attacks increased by 75.60\% \cite{Mustapha-Elsevier-2023}.
According to Markets\&Markets\footnote{\url{https://www.marketsandmarkets.com/Market-Reports/ddos-protection-mitigation-market-111952874.html}},
the global DDoS security and protection market size was valued at USD 3.9 billion in 2022 and is expected to reach USD 10.39 billion by 2030, with a Compound Annual Growth Rate (CAGR) of 12.3\% from 2025 to 2030. Cloud-based DDoS security and protection services can effectively handle volumetric DDoS attacks, as well as layer 3 and 7 attacks. Thus, to optimize operations and costs, companies are rapidly adopting cloud-based anti-DDoS solutions. At the same time, due to the increasing adoption of 5G technology, a survey by the American mobile operator A10 showed that 63\% of respondents believe that advanced DDoS security and protection solutions are needed to protect 5G networks.
Therefore, we consider DDoS detection and mitigation
an important global topic
where modern AI techniques are necessary to fight off attacks,
while, at the same time, helping the victim in maintaining a functioning infrastructure.

In this work, we use
machine learning and artificial intelligence interchangeably
to describe learning-based methods.
Where appropriate, we will use the terms shallow and deep learning 
to separate classical machine learning methods from deep neural networks.
We follow the classification of DDoS attacks according to the Cybersecurity and Infrastructure Security Agency guide provided by the Federal Bureau of Investigation~\cite{CISA-FBI-2024}.
There,
DDoS categories are separated into
volumetric~\cite{Najafimehr-ER-2023},
protocol~\cite{Malliga-ITC-2022, Najafimehr-ER-2023},
and application~\cite{Tripathi-CSUR-2021, Malliga-ITC-2022, Najafimehr-ER-2023} level attacks~\cite{Singh-IJSWIS-2022}.
Volumetric attacks are most common,
often involving bot networks that bombard the victim with a large number of connections that do not have to maintain state,
like UDP packets, for example.
Protocol attacks are more sophisticated,
as they exploit defects found in specific protocol implementations
to employ both network floods,
but also a computational burden on the target.
Reflection and amplification~\cite{Nuiaa-IJASEIT-2022, Najafimehr-ER-2023}
are commonly-used DDoS techniques
meant to widen volumetric and protocol attacks.
Application attacks target high-level services, such as web applications and database management systems,
and mostly inflict large computational loads on the victims.

DDoS defenses are of special interest for
IoT networks
which are more susceptible
and represent a large part of the bot infrastructure mounted 
during volumetric and protocol attacks~\cite{Kadri-IT-2024, Pakmehr-CC-2024}.
SDN
is another particular context where anti-DDoS methodologies
were studied~\cite{Musa-IEEEAccess-2024, Su-E-2024}.
As these represent dynamic network architectures
often used in cloud environments,
their configuration usually includes recent DDoS defenses
which take advantage of the known software-defined topology.

AI-based detection is often treated in the literature as a binary classification task~\cite{Salahuddin-TNSM-2021,Rios-ACMSIGAPP-2024}
performed on
available datasets~\cite{Sharafaldin-ICCST-2019,Sharafaldin-ICISSP-2018,Alzahrani-SRPIS-2018}
containing raw labeled traffic recordings
of normal and attack traffic.
The attacks are usually mixed~\cite{Kayacik-CPST-2005,Moustafa-MilCIS-2015}
as the datasets are designed for IDS~\cite{Sharafaldin-ICISSP-2018}
or IoT networks~\cite{Koroniotis-ELSFGCS-2019,Alsaedi-ACCESS-2020},
and not specifically for the DDoS case.
That is why, among the datasets that do contain DDoS attacks,
most focus on volumetric floods or web application attacks.
Recent trends show an interest in the use of
adversarial training
and adversarial attacks and examples~\cite{Alatwi-ARXIV-2021,He-IEEE-2023}
as a means to provide more robust detection models.

Once detected,
the next critical step for anti-DDoS methods is attack mitigation~\cite{Alatwi-ARXIV-2021}.
Mitigation represents the set of measures taken to block the attack and permit normal traffic to pass through.
These often resume to manual firewall rules, custom made for the current attack.
Unfortunately,
AI literature on the topic is limited 
to 
DTs~\cite{Zadnik-IIS-2023,Coscia-JISAS-2024}
and
a few specific cases,
such as DNS Floods~\cite{Wang-KSII-2012,Ballani-ACM-2008}.
Only recently, we started to see an interest in specializing
deep learning methods in general,
and 
LLM
in particular,
to automatically generate firewall rules~\cite{Louro-ARS-2024,Wang-APNet-2024,Yin-ARXIV-2024}.

While by its very nature the DDoS attack is difficult to topple 
due to its distributed, concentrated and bandwidth depletion nature,
existing detection and mitigation approaches come with their own challenges.
Our survey will go into detail about each of them in the following sections,
still we briefly highlight here the most pressing ones.
First,
existing datasets are mostly over-fitted by current models
that often report 99\% detection rates
while DDoS is still very effective in practice;
we discuss this in Section~\ref{sec:data}
and provide future research directions
for dataset design and cross-dataset generalization.
One might argue that with such high accuracy rates,
there is no need for explainable models
as no auditing is necessary for near perfect models;
still taking into consideration the former challenge
and assuming that it will soon be overcome,
we argue that explainability will be very much required;
we discuss this in Section~\ref{sec:ai-mitigation}
and discuss separately its ethical implications and future directions.
Finally,
we note the scarcity of hybrid models in the literature
where fast static analysis is coupled with separate AI models for each DDoS attack type; this would provide faster detection and mitigation times,
with a decreased number of false positives (due to the static analysis),
an improved accuracy (due to the AI component)
and a lower environmental footprint (due to lower energy consumption) --
we address this across multiple sections and propose future directions.

\noindent\textbf{Contributions.}
In this paper, we present a clear and structured survey
covering all attack categories and attack types,
with a special focus on
their detection and mitigation through machine learning methods.
Although related surveys have touched on these issues (as shown in Section 2), we believe that none of them have considered 
a unified perspective centered around AI methodology.
As such,
we would like to underline the following important contributions
that our study provides:
\begin{enumerate}
    \item \textit{Taxonomy.}
    (a) We provide a thorough hierarchical grouping of the surveyed research papers from the perspective of their main contribution;
    (b) we identify four different attack categories in an attempt to settle existing overlaps and ambiguity found in the literature, especially between protocol and application attacks;
    (c) we introduce an automatic taxonomy based on an agglomerative clustering algorithm, which we discuss and compare;
    
    \item \textit{Data format.} 
    Our work makes a point of discussing and analyzing upfront
    data formatting and preprocessing options
    for the machine learning algorithms;
    starting from the raw network traffic,
    we discuss data representation as
    topological graph data for traffic structure information,
    time series data for aggregate traffic evolution,
    and tabular data for individual packets and coalesced flow information;
    
    \item \textit{Detection.} (a) We provide an in-depth discussion for each of the four attack categories
    including all attack types within;
    (b) we include an important discussion and analysis regarding the time delay between the moment that an attack is mounted and the moment when it is detected and then mitigated;

    \item \textit{Mitigation.} 
    A novelty in our work, due to very recent active research in this new direction, represents the survey of papers dealing with mitigation,
    after the DDoS attack is detected,
    where mitigation is achieved through AI methods
    whose task is to emit efficient firewall blocking rules;

    \item \textit{AI-generated DDoS traffic.}
    Another very recent direction in this field,
    which we survey in our work, are the topics of
    (a) adversarial training
    and
    (b) adversarial examples and attacks --
    these generative AI techniques can help to improve detection and mitigation;

    \item \textit{Research directions.}
    We identify and provide clear and direct ways of 
    improving the field of DDoS detection and mitigation
    using AI;
    although the literature consists of a vast amount of work on this topic,
    we notice a lack of specialized anti-DDoS machine learning methods,
    while
    practical tasks, such as minimizing detection delays together with providing efficient mitigation techniques,
    are just beginning to be explored.
    
\end{enumerate}
\begin{table}
\centering
\tiny
\begin{minipage}[b]{0.5\textwidth}\begin{tabular}{ll}
Acronym & Description \\
\hline
AS & Autonomous System \\
BGP & Border Gateway Protocol \\
DDoS & Distributed Denial of Service \\
DoS & Denial of Service \\
DrDoS & Distributed reflection DoS \\
HDFD & HTTP DDoS Flooding Defender \\
IDF & Inverse Document Frequency \\
IDS & Intrusion Detection System \\
IoT & Internet of Things  \\
MAN & Metropolitan Area Networks \\
PoD & Ping of Death \\
RA-DDoS & Reflection and amplification DDoS \\
R.U.D.Y. & R U Dead Yet attack\\
SDN & Software Defined Network \\
SIP & Session Initiation Protocol \\
TF & Term Frequency \\
TF-IDF & TF-Inverse Document Frequency \\
ToS & Type of Service \\
TTL & Time To Live \\
XDP & eXpress DataPath \\
\hline
AIC & Akaike Information Criterion~\cite{Bozdogan-JMP-2000} \\
ADASYN & Adaptive Synthetic Sampling~\cite{He-IJCNN-2018} \\
ANOVA & Analysis of variance~\cite{Pham-ASE-2020} \\
BiGAN & Bidirectional GAN~\cite{Zhang-CoNLL-2018} \\
CNN & Convolutional Neural Networks~\cite{Ibrahim-CSUR-2023} \\
CoD & Constraint-of-Deviation~\cite{Liu-CHI-2024} \\
CoT & Chain-of-Thought~\cite{Yu-ARXIV-2023} \\
CTGAN & Conditional GAN~\cite{Bourou-ARXIV-2024} \\
DAGMM & Deep Autoencoding GMM~\cite{Zong-ICLR-2018} \\
DT & Decision Trees~\cite{Costa-AIR-2023} \\
\hline\end{tabular}\end{minipage}\begin{minipage}[b]{0.5\textwidth}\begin{tabular}{|ll} Acronym & Description \\\hline
EAD & Elastic-net Attacks to DNNs~\cite{Chen-AAAI-2018} \\
ED & Euclidean distance \\
EWMA & Exponentially Weighted Moving Average~\cite{Sukparungsee-PLOS-2020} \\
FCM & Fuzzy C-means~\cite{Nayak-CIDM-2015} \\
FGSM & Fast Gradient Sign Method~\cite{Safaryan-ICML-2021} \\
FL & Fuzzy Logic~\cite{Khan-CSUR-2022} \\
GA & Genetic Attack~\cite{Huang-IEEE-2020} \\
GAN & Generative Adversarial Networks~\cite{Zhang-CSUR-2024} \\
GMM & Gaussian Mixtures Model~\cite{Bouguila-SPRINGER-2020}\\
GNB & Gaussian Naive-Bayes~\cite{Reddy-BRGPMLA-2022} \\
GRU & Gated Recurrent Units~\cite{Chung-NIPS-2014} \\
IF & Isolation Forest~\cite{Xu-TKDE-2023} \\
JSMA & Jacobian-based Saliency Map Attack~\cite{Wiyatno-ARXIV-2018} \\
k-NN & k-Nearest Neighbors~\cite{Cunningham-CSUR-2021} \\
LGBM & Light Gradient Boosting Machine~\cite{Fan-AWM-2019} \\
LLM & Large Language Model~\cite{Zheng-CSUR-2025} \\
LoRA & Low-Rank Adaptation~\cite{Wong-CSUR-2024} \\
LSTM & Long Short-Term Memory~\cite{Van-AIR-2020} \\
MLP & Multi-Layer Perceptron \\
MNB & Multinomial Naive-Bayes~\cite{Reddy-BRGPMLA-2022} \\
NB & Naive Bayes~\cite{Reddy-BRGPMLA-2022} \\
OC-SVM & One-Class SVM~\cite{Seliya-JBD-2021} \\
PEFT & Parameter Efficient Fine-Tuning~\cite{Han-ARXIV-2024} \\
PWPSA & Probability Weighted Packet Saliency Attack~\cite{Huang-IEEE-2020} \\
RF & Random Forest \\
SMOTE & Synthetic Minority Over-sampling Technique~\cite{Chawla-JAIR-2002}  \\
SVM & Support Vector Machine \\
TVAE & Tabular Variational Autoencoder~\cite{Xu-NIPS-2019} \\
VAE & Variational Autoencoder~\cite{Ghojogh-ARXIV-2021} \\
XGB & Extreme Gradient Boosting \\
\hline
\end{tabular}\end{minipage}
\caption{List of terms and acronyms used in the survey.
    Grouped at the top are network and attack terms in alphabetical order,
    followed at the bottom
    by artificial intelligence nomenclature.
    For the interested reader,
    we added references to recent surveys and papers
    that discuss the topic further.}
\label{tab:nomenclature}
\end{table}
\noindent\textbf{Outline.}
The survey is organized in multiple sections
with the list of terms and abbreviations collected in Table~\ref{tab:nomenclature}.
Section~\ref{sec:surveys} discusses existing surveys
by comparing our work with them and motivating the need for the current manuscript.
In Section~\ref{sec:taxonomy},
we carefully define and describe the attack categories
and attack types within each category
in order to group the selected papers from this survey
through the use of both manual
and automatic taxonomies.
Section~\ref{sec:data}
discusses available datasets
and,
more importantly for machine learning algorithms,
the data formats and preprocessing techniques
that are used by recent learning algorithms.
Next,
we delve into the details regarding
DDoS detection, in Section~\ref{sec:detection},
and AI-based mitigation, in Section~\ref{sec:ai-mitigation}.
The detection section includes all the attack categories 
included in the taxonomy
(volumetric, protocol and application),
but also a dedicated subsection for reflection and amplification techniques.
In Section~\ref{sec:ai-traffic},
we discuss how to further improve existing algorithms, models and datasets
through adversarial training and adversarial examples.
Finally,
in Section~\ref{sec:conclusion},
we discuss future research opportunities
for the directions described above
along with conclusion after our intensive survey activity.

\section{Related Surveys}
\label{sec:surveys}
\begin{table}
    \centering
    \setlength\tabcolsep{0.4em}
    \tiny{
    \begin{tabular}{r|cc|cccc|ccc|cc|c|l}
    \multirow{6}*{Survey} 
    & \multicolumn{2}{c|}{Volumetric} 
    & \multicolumn{4}{c|}{Protocol}
    & \multicolumn{3}{c|}{Application} 
    & \multicolumn{2}{c|}{Generative} 
    & \multirow{6}*{\rotatebox{90}{Mitigation}}
    & \multirow{6}*{Limitations} \\
    & \rotatebox{90}{Flood} & \rotatebox{90}{Reflection}
    & \rotatebox{90}{TCP} & \rotatebox{90}{Ping} & \rotatebox{90}{BGP} & \rotatebox{90}{Smurf}
    & \rotatebox{90}{HTTP} & \rotatebox{90}{DNS} & \rotatebox{90}{Slow}
    & \rotatebox{90}{Training} & \rotatebox{90}{Adversarial} && \\
    \hline
    \textbf{Ours} & \ccmark & \ccmark & \ccmark & \ccmark & \ccmark & \ccmark & \ccmark & \ccmark & \ccmark & \ccmark & \ccmark & \ccmark \\
    \hline
    Malliga et al. \cite{Malliga-ITC-2022}
    &  & 
    & \ccmark & \ccmark & &
    & \ccmark &  & 
    & \ccmark &
    & & Missing volumetric\\
    Najafimehr et al. \cite{Najafimehr-ER-2023}
    & \cmark & \cmark
    & \cmark & \cmark && \cmark
    & \cmark & \cmark & \cmark
    &&
    & &Missing generative\\
    Tripathi et al. \cite{Tripathi-CSUR-2021}
    & &
    &  &  &&
    & \ccmark & \ccmark & \ccmark
    &&
    && Application only\\
    \hline
    Singh et al. \cite{Singh-IJSWIS-2022}
    &&
    & \cmark &&& \cmark
    & \cmark && \cmark
    &&
    && Attack only\\
    Nuiaa et al. \cite{Nuiaa-IJASEIT-2022}
    && \cmark
    &&&&
    && \ccmark & \ccmark
    &&
    & & DRDoS and Attack only\\
    \hline
    Kadri et al. \cite{Kadri-IT-2024} 
    & \ccmark &
    & \ccmark & & &
    &&&
    & \cmark &
    & & Missing application, IoT only\\
    Pakmehr et al. \cite{Pakmehr-CC-2024}
    & \cmark & 
    & \cmark & \cmark & & \cmark
    & \cmark & \cmark & \cmark
    & \cmark &
    & & IoT only\\
    \hline
    Musa et al. \cite{Musa-IEEEAccess-2024}
    & \ccmark &
    & \ccmark & \ccmark &&
    & \ccmark & \cmark & \ccmark
    && \ccmark
    & & SDN only \\
    Su et al. \cite{Su-E-2024}
    & \ccmark &
    & \ccmark && \cmark &
    &&&
    &&
    & & Missing application, SDN only\\
    \hline
    Alatwi et al.~\cite{Alatwi-ARXIV-2021}
    &&
    & \ccmark &&&
    &&&
    & \cmark & \ccmark
    & \cmark & Protocol only, Generative only \\
    He et al.~\cite{He-IEEE-2023} 
    &&
    &&&&
    & \cmark &&
    & \ccmark & \ccmark
    && Application only, Generative only\\
    \hline
    \end{tabular}
    }
    \caption{Comparison of covered AI-based DDoS detection and mitigation topics in related surveys;
    \ccmark means AI detection or mitigation is directly addressed in the survey,
    \cmark means the topic is mentioned but treated in bulk with others or indirectly covered.}
    \label{tab:surveys}
\end{table}

While existing literature contains plenty survey papers on the DDoS topic,
most are centered on the attacks and few on the detection and mitigation tasks.
Out of the latter, even fewer include learning based approaches,
as can be seen in Table~\ref{tab:surveys}:
only the three papers at the top employ a general study of the field,
while the ones at the bottom focus on specific tasks and testbeds.

Anti-DDoS surveys studying the general problem
from a machine learning perspective
focus exclusively on the detection task, without discussing learning-based mitigations.
Also,
these generic surveys discuss only a share of
existing DDoS categories,
mostly involving a subset of
volumetric, protocol or application attacks,
but never all of them together.
Malliga et al.~\cite{Malliga-ITC-2022} cover mainly protocol-based attacks
consisting of 
a table of 66 research papers on the use of deep machine learning methods for DDoS detection
and
a table of 12 of the most popular datasets used in simulations.
The article gives a good tabular description of the field, but does not detail the methods analyzed and does not discuss the methods of generating DDoS attacks.
Volumetric attack detection
and mitigation strategies
are completely missing.
Interested readers can continue with application-level attacks
from the work of
Tripathi et al.~\cite{Tripathi-CSUR-2021}, who
provide a thorough survey regarding
exclusively these types of attacks and detection mechanisms,
while also including existing turnkey solutions from the industry.
Unfortunately, the authors tackle
few existing learning-based detection techniques, most of which are shallow.
Further on,
Najafimehr et al.~\cite{Najafimehr-ER-2023} present a detailed taxonomy of DDoS attack detection methods
which includes the missing volumetric attack types.
The survey is centered around pairings between
general detection methods and existing databases 
with a bias towards shallow learning approaches.
Detecting specific attacks is not treated as a separate topic
nor the particularities of each task;
the algorithms are expected to detect with high accuracy
any type of DDoS attack present in the dataset.

It is worth mentioning that
some of the DDoS attack surveys
also include a small section dedicated to learning-based detection of these attacks.
We mention here the work of
Singh et al.~\cite{Singh-IJSWIS-2022},
which enumerates the list of machine and deep learning papers dealing with some protocol and application based attacks,
and that of
Nuiaa et al.~\cite{Nuiaa-IJASEIT-2022},
which analyzes solely DDoS attacks against the DNS protocol
focusing on distributed methods with reflection (and amplification)
that includes a small discussion about
using AI for the attack detection task.

Moving on towards specialized surveys,
we note that
reflection, amplification, mitigation and attack generation techniques are not discussed in these works.
The IoT domain
is presented by 
Kadri et al.~\cite{Kadri-IT-2024},
who discuss floods and TCP-based attacks
analyzing detection and validation methods used in the literature.
The paper talks little 
about the performance of these methods in simulations and practical applications.
The work of
Pakmehr et al.~\cite{Pakmehr-CC-2024} 
picks-up on this and
presents several tables
with advantages and disadvantages for shallow and deep learning methods;
these tables are not very well connected to the survey text or the particularities of the attacks exploited by the detection methods.
The paper covers a wider area of attacks,
including a brief mention of modern GAN-based learning strategies.
Nonetheless, techniques for generating AI-based DDoS attacks are not discussed.

SDNs are a hot topic in the DDoS field,
and indeed, we also find two attack detection surveys specializing on it.
The work of Musa et al.~\cite{Musa-IEEEAccess-2024}
tackles the detection task
with an in-depth and wide coverage across the DDoS attack classes
describing recent learning methods (both classical and deep)
including Adversarial AI approaches.
The article laments the simplistic scenarios discussed in the research articles and the need to test more complex scenarios that integrate with already existing security systems.
While the authors provide a thorough review of the SDN literature on the topic,
the survey does not provide insights and connections between the selected papers
employing instead a list of abstracts approach.
The mitigation task is not covered here nor in other SDN surveys.
Even though
Su et al.~\cite{Su-E-2024} conduct a study of most mitigation methods for SDN DDoS attacks,
learning methods are used solely in the detection process.
These range from statistical to shallow and deep learning methods together with hybrid variants.


While significant progress has been made in the field, there are still some major obstacles, especially regarding the quality of the datasets and the adaptability of the defense systems. 
Even though not DDoS-centric,
Alatwi et al.~\cite{Alatwi-ARXIV-2021} focus on three main issues: classifying studies from the existing literature that deal with generating adversarial examples, evaluating deep learning-based detection systems, and proposing defense mechanisms against such attacks. The authors state that approaches such as adversarial training, although effective in areas like computer vision, fail to perform as well in network IDS because of the complex structure of traffic characteristics. While making some interesting observations, the survey of Alatwi et al.~\cite{Alatwi-ARXIV-2021} covers a narrow domain, adversarial learning,
dealing with a generic problem,
network intrusion detection.

Supported by the views in \cite{Alatwi-ARXIV-2021}, He et al.~\cite{He-IEEE-2023} talk about the vulnerabilities and limitations of machine learning-based methods in network IDS in the context of adversarial attacks. As noted by previous research, these systems are vulnerable to attacks that manipulate network traffic to avoid detection. Still, He et al.~\cite{He-IEEE-2023} argue that many of these attacks lack adaptability to the data structure of the network. They conclude that the literature still needs more representative and diverse datasets, along with more robust defense mechanisms such as adversarial training and feature reduction strategies. The survey of He et al.~\cite{He-IEEE-2023} is focused on a specific area related to adversarial attack generation. 

The DDoS literature contains several comprehensive surveys, most are focused on attack taxonomy, and a smaller subset focus on learning-based approaches.
Among these,
general AI-based surveys focus almost exclusively on detection
specializing on subsets of DDoS categories
rather than volumetric, protocol, and application-layer all-together.
These surveys often emphasize results obtained on benchmark datasets
more than a critical analysis of techniques,
attack-generation mechanisms,
or custom mitigation strategies.
As a result,
while specialized and in-depth,
the existing literature remains quite fragmented.
We position our survey
as a unified AI-based synthesis of the field
that provides a single coherent framework
jointly addressing the diversity of DDoS attacks (Section~\ref{sec:taxonomy}),
the realism of datasets coupled with the AI-based evaluation and generalization capabilities (Section~\ref{sec:data}),
detection across all attack types (Section~\ref{sec:detection}),
AI-based mitigation (Section~\ref{sec:ai-traffic}),
attack generation (Section~\ref{sec:ai-mitigation}),
together with a discussion regarding ethical considerations and societal impact (Section~\ref{sec:ethics}).
In the next sections of this survey, we will endeavor to address these topics one by one.

\section{Taxonomies of DDoS Attacks and AI-Based Detection/Mitigation Solutions}
\label{sec:taxonomy}

In this section,
we start by describing our literature selection process
for the different taxonomies involving DDoS attacks
and their AI-based mitigation solutions.

We surveyed research articles from 2016 to 2026 that studied DDoS detection and mitigation techniques that process network data and make decisions with Artificial Intelligence.
This includes both shallow and recent neural network-based algorithms.
Rarely and only where necessary, we went beyond 2016 to retrieve relevant and well-established papers in the field that mostly deal with
taxonomies, datasets, or network tools.
We focus on works that go beyond simply applying existing AI methods or present datasets and report the results, and instead promote papers that take advantage of the specific network structure, as in Section~\ref{sec:data}, or that integrate the detection in a more complex setup, partially involve attack classification, mitigation or integration in Intrusion Detection Systems (IDS).



\begin{figure}
  \centering
     \includegraphics[scale=0.7]{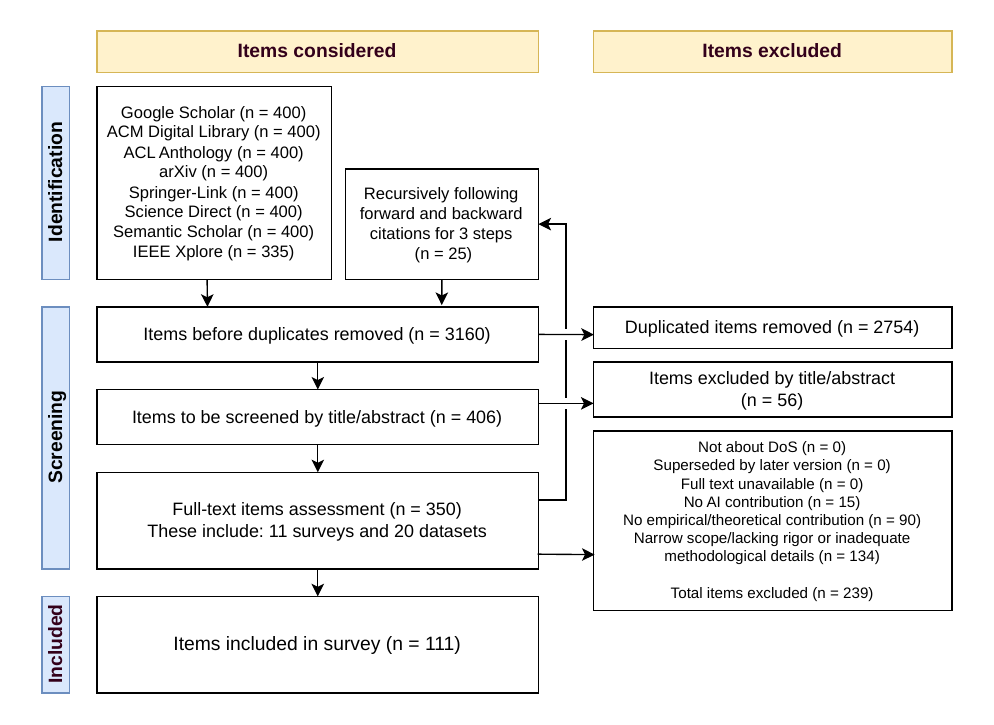}
     \caption{PRISMA-like diagram of the literature selection process.}
     \label{fig:PRISMA}
\end{figure}

We queried a broad set of digital libraries in order to ensure representative coverage for the initial pool.
The initial pool was gathered by searching for \textit{AI DDoS}, \textit{Denial of Service} on ACM Digital Library, ACL Anthology, SpringerLink, ScienceDirect, Google Scholar, arXiv, Semantic Scholar and IEEE Xplore.
Only the first 400 entries were considered, in the order of the default filters for each one of the platforms. Following this step, we filtered the relevant articles only, based on the title and abstract. A full list of the discarded papers was not maintained. After the full-text review of each relevant article, the missing relevant citations and the backward references\footnote{Found using Google Scholar} were added to the pool. A diagram outlining the process can be found in Figure \ref{fig:PRISMA}.

We then examined the main contributions of the remaining papers
to assess whether they were a suitable fit for our survey.
Ultimately, a study was included in our survey if it matched any of the
following criteria:
(a) analyzes and adapts AI-based methods in the context of DDoS detection problems, 
(b) discusses and proposes AI-based DDoS mitigation solutions,
(c) clearly establishes attack taxonomy, or
(d) introduces a new DDoS detection dataset.
A comprehensive list of excluded studies was not kept.

For the proposed dendrogram in Figure~\ref{fig:dendro_ward} from Section~\ref{sec:taxonomy-tfidf},
we further restricted the selected papers to the ones that
focus on specific DDoS detection, mitigation, or adversarial topics.
We eliminated papers that treated DDoS only as a particular example (but mainly focused on other technical research aspects),
and papers representing surveys, datasets, taxonomies, or network aspects.
In the end, we removed singleton clusters,
papers that did not match any other similar recent research papers.
In the interest of transparency and reproducible research,
we make the automatic taxonomy algorithm publicly available
together with the full and restricted set of papers,
used in Figure~\ref{fig:dendro_ward},
at
\url{https://codeberg.org/pirofti/anti-ddos-with-ai-survey}.

To avoid potential ambiguities, we first endeavor to introduce a set of clear definitions for the DDoS attack categories 
that are meant to easily and unambiguously separate existing attack types.
Next,
we provide a manual hierarchical grouping (taxonomy) of existing research on AI-based DDoS detection methods,
together with an automatic clustering-based taxonomy,
and compare the two.

\subsection{DDoS attack classification}

\begin{figure}
  \centering
     \includegraphics[scale=0.55]{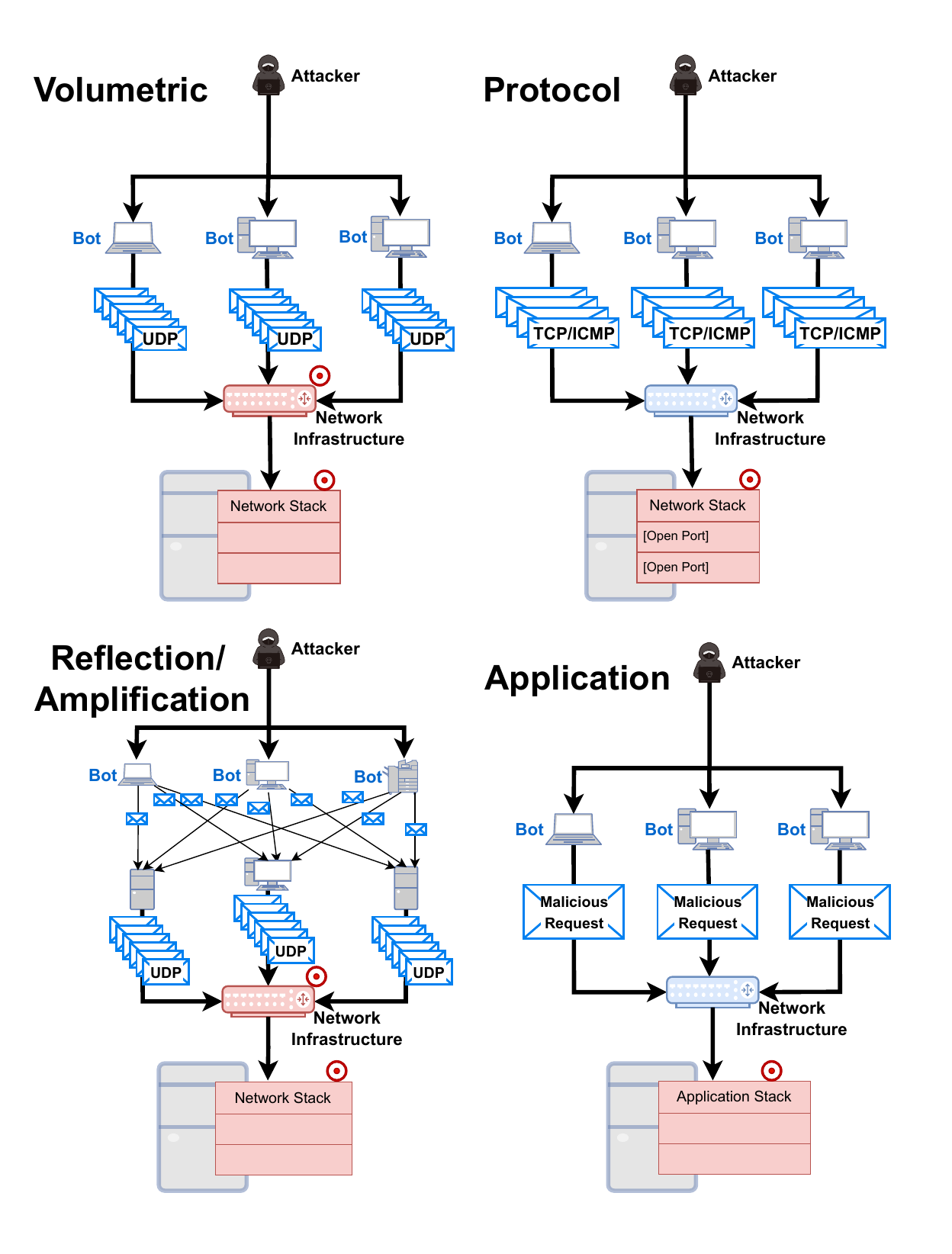}
     \caption{Taxonomy of the four major categories of DDoS attacks described in separate diagrams. In each diagram, we depict an attacker who has control over networks of bot machines and targets a particular victim. Packets involved in the attacks bear the inscriptions of the typically used protocols, while on the victim side, we highlight in red and mark with bullseye the network resources that are the focus of the attacks: the network infrastructure such as communication hardware, the network software stack of the victim operating system, and the application software stack running user services.}
     \label{fig:DDoSAttacks}
\end{figure}
\begin{table}
\centering
\tiny
\begin{talltblr}[caption={A description of the most popular DDoS approaches from each of the four attack categories. We highlight the main attack target (network infrastructure and/or computational resources) and concisely describe the attack.},
label={tbl:DDoSattacks}]
{
  row{1} = {c},
  cell{2}{1} = {r=3}{c},
  cell{2}{3} = {r=3}{},
  cell{5}{1} = {r=5}{c},
  cell{6}{3} = {r=4}{},
  cell{10}{1} = {r=2}{c},
  cell{10}{3} = {r=2}{},
  cell{12}{1} = {r=7}{c},
  cell{13}{3} = {r=6}{},
  vline{2-4} = {1-2,5,10,12}{},
  vline{2-4} = {1-18}{},
  vline{3} = {3-4,7-9,11,14-18}{},
  vline{3-4} = {6,13}{},
  hline{2,5,10,12,19} = {-}{},
  hline{3-4,7-9,11,14-18} = {2,4}{},
  hline{6,13} = {2-4}{},
}
                                          & Attack name                                               & Target                & Brief description                                                                                                                                                         \\
\begin{sideways}Volumetric\end{sideways}  & {UDP flood\\(DNS, NTP, SSDP, IPsec,\\QOTD, SNMP, QUIC)}   & Network               & {Large number of UDP packets, sent via high-level\\protocols which are based on UDP}                                                                                      \\
                                          & ICMP fragmentation flood                                  &                       & Malformed  ICMP packets                                                                                                                                                   \\
                                          & Ping flood (IP/ICMP)                                      &                       & Large number of ICMP packets                                                                                                                                              \\
\begin{sideways}Protocol\end{sideways}    & {TCP flood\\(SYN/ACK/RST/FIN)}                            & {Network/\\Compute}   & {Large number of TCP (SYN/ACK/RST) packages,\\connections not fully established}                                                                                          \\
                                          & Ping of  Death (IP/ICMP)                                  & Network               & Malformed ICMP packet(s)                                                                                                                                                  \\
                                          & Smurf  DDoS (ICMP flood)                                  &                       & Large number of spoofed ICMP packets                                                                                                                                      \\
                                          & BGP flood                                                 &                       & {Malformed Border  Gateway Protocol packets,~\\disrupt the routing tables}                                                                                                \\
                                          & {DHCP Starvation\\(IP + MAC)}                             &                       & {Large number of requests from spoofed\\MAC addresses which consume the DHCP IP pool}                                                                                     \\
\begin{sideways}Reflection/Amplification\end{sideways}            & {UDP flood\\(DNS, NTP, SSDP, QOTD,\\RPC, NetBIOS, CLDAP)} & {Network/\\ Compute } & {Large number of UDP packets, sent via high-level\\protocols which are based on UDP, redirected from \\multiple vulnerable machines or bot networks,\\manyfold amplified} \\
                                          & {TCP flood\\(SYN/ACK/RST/FIN)}                            &                       & {Large number of TCP (SYN/ACK/RST) packages,\\connections not fully established, redirected from \\multiple vulnerable machines or bot networks,\\manyfold amplified}     \\
\begin{sideways}Application\end{sideways} & {HTTP  GET/POST flood\\(TCP flood)}                       & {Network/\\Compute}   & {Large number of concurrent GET/POST\\requests or send large files via HTTP}                                                                                              \\
                                          & HTTP Slowloris/R.U.D.Y.                                   & Compute               & {Large number of concurrent slow or incomplete\\HTTP requests}                                                                                                            \\
                                          & {HTTP  Initiated\\(HTTP Asynchronous)}                    &                       & {Malicious HTTP requests that target web\\application logic/features}                                                                                                     \\
                                          & DNS  Server Query Flood                                   &                       & Large number of legitimate-looking DNS queries                                                                                                                            \\
                                          & {DNS  NXDOMAIN/\\Water torture}                           &                       & Large number of~queries for non-existent domains                                                                                                                          \\
                                          & ReDoS                                                     &                       & {Malicious regex search patterns that\\overload compute resources}                                                                                                        \\
                                          & Database  DDoS                                            &                       & Malicious HTTP requests that strain the DBMS                                                                                                                              
\end{talltblr}
\end{table}

In Figure~\ref{fig:DDoSAttacks}, we provide an overview of the four general categories of DDoS attacks, which we describe individually next.

\noindent The most common attacks in the real-world are volumetric DDoS attacks, in which an attacker coordinates bot networks to send a large number of UDP packets to a target victim. This increased traffic affects network and compute infrastructures, which now have to appropriately handle the increased fake traffic in detriment of traffic from legitimate users.

Next, protocol DDoS attacks exploit weaknesses in protocols themselves (mostly network layer protocols, such as TCP/IP/ICMP) and their software implementations and usually target the compute infrastructure of the victim operating system. This leads to an increase in the consumption of hardware resources to the point where the service becomes unable to handle legitimate requests.

Many of the attacks described in the volumetric and protocol categories can be further enhanced by reflection/amplification DDoS techniques. In this case, the attacker uses a bot network to take advantage of the connectionless nature of UDP to send requests with a spoofed IP address to multiple legitimate UDP-based services. The responses from these so-called reflectors, which are usually also amplified by a certain factor, overwhelm the target victim, further exacerbating the disruption of both network and compute capabilities.

Finally, application DDoS attacks target specific higher-level network services, including HTTP, web applications, and database management systems. In this scenario, an attacker sends a large number of well-crafted malicious requests to a network application that cannot properly handle the requests. This leads to increased consumption of hardware resources that ultimately make the application nonresponsive or outright crash the application, and thus denying the services to legitimate users. In this scenario, the network infrastructure is mostly unaffected and the main target is the victim application stack on top of the operating system.

In Table~\ref{tbl:DDoSattacks}, we provide a list of the most common DDoS attacks, grouped into each of the four main attack categories, together with a brief description and highlighting the main attack target: network infrastructure and/or computational resources.

\subsection{Existing AI-based DDoS detection research}
\begin{figure}
    \centering
    \includegraphics[width=\linewidth]{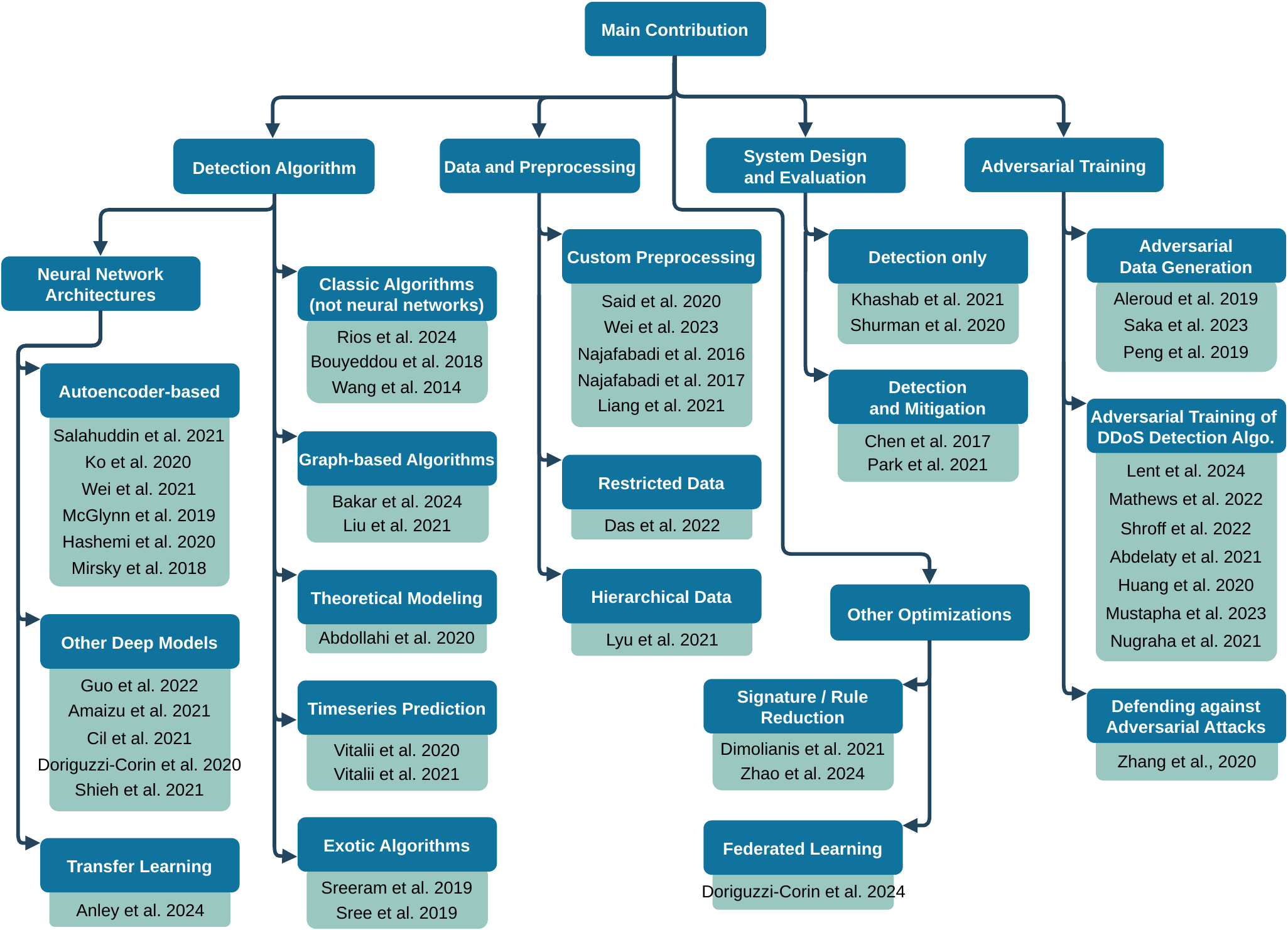}
    \caption{Hierarchical grouping of the surveyed articles according to their main innovative contribution. References are linked to the papers (click on a reference to open the DOI page of the article).}\label{fig:taxonomy-goals}
\end{figure}

In an attempt to help the reader quickly navigate the state-of-the-art landscape
and identify key affinities, 
we group the research papers according to their main innovative contribution, 
e.g., whether they focus on a particular detection algorithm, a new data preprocessing method, or building and evaluating a full DDoS mitigation system, etc. 
\hlight{R3-C5}{For the manual grouping in Figure \ref{fig:taxonomy-goals}, we assigned each paper according to its perceived primary claimed novelty. We first identified the main task emphasized by the paper, e.g., detection, mitigation, adversarial generation/training, or system-level evaluation. Within that task, we then assigned the paper to exactly one dominant category. When a study contributed along multiple axes, we used the title, abstract, methodological core, and experimental emphasis to determine the main contribution and selected a single category to avoid double-counting. We stress that this manual taxonomy is intentionally reader-oriented, whereas the automatic taxonomy from Section \ref{sec:taxonomy-tfidf} provides a complementary data-driven organization of the same literature.}


As illustrated in Figure~\ref{fig:taxonomy-goals}, the largest group focuses on the actual algorithm for detecting DDoS attacks.
The majority of the proposed approaches rely on deep neural networks,
the autoencoder being a particularly popular architecture,
thanks to its ability to detect deviations and abnormal patterns.
However, several other custom architectures have been proposed, e.g., using 
bidirectional GRU layers with self-attention \cite{Guo-SCN-2022}. 
Of particular interest is study of Anley et al.~\cite{Anley-CS-2024}, which also investigates how well
the detection results transfer to other datasets than the ones used for training.
Besides neural networks, various other algorithms are investigated, including classic supervised solutions (e.g., RFs), graph-based kernels \cite{Liu-JNCA-2021}, various modelings of traffic patterns, and even bio-inspired heuristic 
classification algorithms.

The second group contains papers with novelties related to data and preprocessing, 
ahead of the actual classification stage. 
A typical example of these custom methods is the approach of Wei et al.~\cite{Wei-arXiv-2023}, which uses LSTM
coupled with an autoencoder to extract time-varying features of the data, 
identifying anomalies using the autoencoder reconstruction error.
Another example is the method of Das et al.~\cite{Das-CSR-2022}, which restricts the data used in the detection process only to port statistics, without any flow-related data, unlike the vast majority of related approaches.

Several articles, grouped under the ``Other Optimizations'' label, focus on collateral optimizations which may be highly relevant
for the efficiency of DDoS solutions in practice. 
Thus, Dimolianis et al.~\cite{Dimolianis-ICIN-2021} and Zhao et al.~\cite{Zhao-CS-2024} consider the problem of consolidating attack signatures or mitigation rules, to detect and block several types of attacks simultaneously, with few rules. 
This also increases the robustness against new, unseen variants of the attacks.
An adaptive federated learning approach is proposed in \cite{DoriguzziCorin-CS-2024}, facilitating cooperation between entities while avoiding the risk of exposing sensitive information.

The ``System Design and Evaluation'' group contains papers that, without focusing on a single particular aspect, 
are relevant to evaluations of DDoS detection pipelines. 
We differentiate here between articles focusing only on the detection, e.g.~\cite{Khashab-Netsoft-2021}, and articles that also consider the mitigation process, e.g.~\cite{Park-IEICETIS-2021}.

The last group gathers articles involving adversarial data and training, 
as a way of increasing the robustness of anti-DDoS solutions.
Since these articles typically use specific algorithms, rarely encountered in other articles (e.g., GAN architectures), 
we group all of them into a separate category.
We further subgroup the articles according to whether they focus primarily on the actual data generation or the adversarial training process, with a separate note for the study of Zhang et al.~\cite{Zhang-ACMSIGSAC-2020}, who investigate strategies against adversarial attacks 
(e.g.~ensemble voting).

While the majority of the reviewed works apply generic ML/DL models on the existing datasets, a small number of papers introduce new algorithms that are specifically designed to take advantage of the specificities of DDoS attacks. For example, Liu et al~\cite{Liu-JNCA-2021} exploit the network structure changes induced by DDoS attacks and Lyu et al.~\cite{Lyu-TNSM-2021} introduce a hierarchical graph structure, specialized in detecting DNS-based DDoS attacks.

\begin{figure}[H]
    \centering
    \includegraphics[width=.98\linewidth]{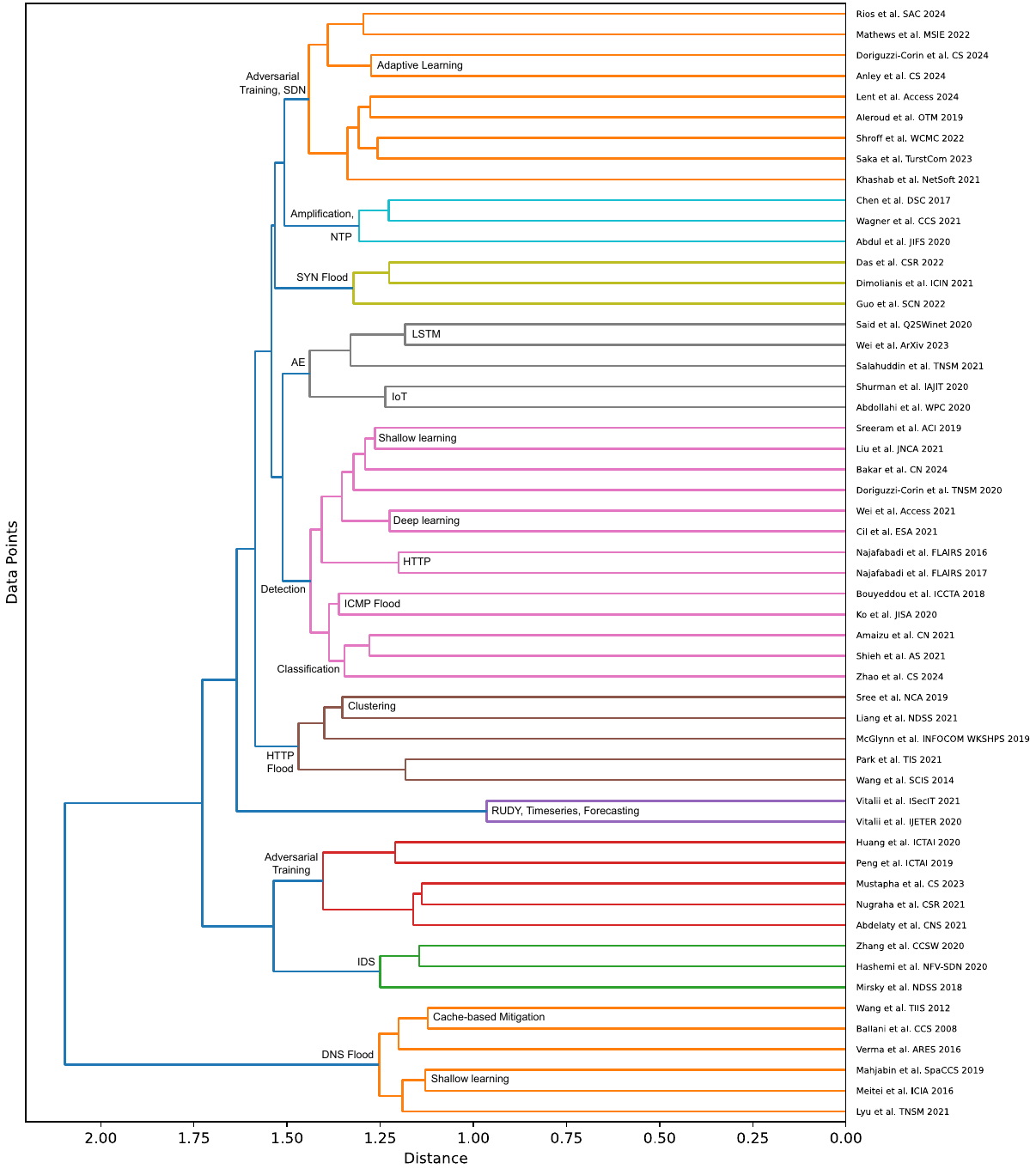}
    \caption{A hierarchical clustering of the surveyed articles based on Ward's linkage. Each article is represented through a TF-IDF vector computed from the concatenated title and abstract of the respective article. The distance represented on the horizontal axis is computed via Eq.~\eqref{eq_ward_linkage}. The dendrogram is manually annotated to indicate meaningful groups of related papers. Best viewed in color.}
    \label{fig:dendro_ward}
\end{figure}

\subsection{Manual vs. automatic taxonomies of AI-based DDoS detection methods}\label{sec:taxonomy-tfidf}

\subsubsection{Hierarchical clustering of articles}

The dendrogram presented in Figure \ref{fig:dendro_ward} is generated by applying an agglomerative clustering algorithm based on Ward's linkage over the TF-IDF representation of article titles and abstracts. \hlight{R1-C3} {All the information for reproducing the figure are present in the associated repository.}

The TF-IDF representation is an extension of the simple bag-of-words model, which combines  two components: term frequency (TF) and inverse document frequency (IDF). TF measures how often a word appears in a document, while IDF demotes common words by assigning lower importance to terms that appear in many documents. The result is a weighted vector representation that emphasizes unique and meaningful words for distinguishing documents.

Agglomerative clustering based on Ward's linkage is a hierarchical clustering algorithm that groups data points by iteratively merging clusters to minimize the total within-cluster variance. It starts with each TF-IDF vector as its own cluster and merges pairs of clusters, step by step. Ward's method chooses the pair of clusters to merge based on the smallest increase in variance within all clusters.

In the following, we denote
$C_i$ as cluster $i$,
with mean vector $\mu_{C_i}$,
and cardinality $|C_i|$.
Let $x$ be a vector with elements $x_k$,
then $\| x\|_2 = \sqrt{\sum_k x_k^2}$ is its $\ell_2$ norm.
The Sum of Squares Error (SSE) of a given cluster $C_i$ is:
\begin{equation}
    \label{eq_cluster_sse}
\mbox{SSE}(C_i)=\sum_{x \in C_i} \| x - \mu_{C_i}\|_2^2.
\end{equation}
Ward's method defines the linkage between two clusters as:
\begin{equation}\label{eq_ward_linkage}
    D(C_i,C_j) = \frac{|C_i|\cdot |C_j|}{|C_i \cup C_j|}\| \mu_{C_i} - \mu_{C_j} \|_2^2 = \mbox{SSE}(C_i \cup C_j ) - \mbox{SSE}(C_i) - \mbox{SSE}(C_j). 
\end{equation}
In the right-hand side of Eq.~$\eqref{eq_ward_linkage}$, the SSE of each individual cluster is subtracted from the SSE of the merged cluster. 


Formally, the optimization criterion used to select the pair of clusters to be merged is defined as:
\begin{equation}\label{eq_cluster_objective}
    \arg\min_{i \neq j}\left\{D( C_i, C_j ) \right\}.
\end{equation}
This approach tends to produce compact clusters and is often effective when the underlying groups are roughly spherical and similarly scaled in Euclidean space. We use the SciPy implementation\footnote{\url{https://docs.scipy.org/doc/scipy/reference/generated/scipy.cluster.hierarchy.ward.html}} of hierarchical clustering based on Ward's linkage.

\hlight{R3-C3}{
\subsubsection{Empirical comparison of hierarchical clustering alternatives}

To motivate the implementation choices for the hierarchical clustering algorithm, we hereby compare alternative linkage \hlight{R1-C2}{criteria (single, complete, average, median and Ward), and two distinct representations (TF-IDF and BERT \cite{Devlin-NAACL-2019} embeddings).}

To automatically measure performance, we first calculate inconsistency statistics on the resulting linkage matrices. For each matrix, we then compute the maximum inconsistency coefficient for each non-singleton cluster and its children. The final evaluation measure is the average of maximum inconsistencies. The corresponding results are reported in Table \ref{tab:hc_results}.

\begin{table}
\centering
\small
\begin{tabular}{r | c c c c c}
Representation & Single & Complete & Average & Median & Ward \\
\hline
TF-IDF  & 0.8119 & 0.5794 & 0.6131 & 0.6938 & \textbf{0.5736}  \\
BERT    & 0.6823 & 0.6071 & 0.6302 & 0.5853 & 0.5804  \\
\end{tabular}
\caption{Average maximum inconsistency scores of linkage matrices obtained with various linkage criteria applied over TF-IDF and BERT representations. Lower values indicate a better clustering. The best score is highlighted in bold.}
\label{tab:hc_results}
\end{table}



In terms of data representation, we observe that TF-IDF yields smaller inconsistency statistics for Ward, complete and average linkage criteria. In terms of linkage criteria, the most competitive options are Ward and complete. For both representations, Ward obtains the smallest average inconsistency. Overall, the reported statistics confirm that our implementation choices, TF-IDF and Ward, produce a more consistent dendrogram than other choices.}

\subsubsection{Manual interpretation of dendrogram}

\hlight{R3-C2}{Clustering itself is an unsupervised learning task, so there are no ground-truth labels to compare the dendrogram against. To interpret the obtained dendrogram,} we manually labeled the resulting clusters after automatically obtaining the taxonomy,
by investigating the grouped papers and
finding the commonality from the point of view of a human expert.
Next,
we briefly walk from
the outer clusters towards the inward labeled clusters
and provide a few comments.
We found that
the first level splits between
DNS, HTTP, and SYN flood attacks
and also creates a clear split between
Adversarial Training in normal and SDN contexts.
Another fine line was the separation between Detection, a large cluster,
and IDS focused research, a niche topic.
Going deeper inside the Detection cluster,
we find clusters separating shallow and deep learning,
and
also a cluster that grouped attack detection and attack classification papers.
There is another shallow learning cluster within DNS Floods
that was accurately separated from the general detection cluster.
In the final clustering level,
the manual labels provide a specialized view of the resulting dendrogram.
For example,
we have DNS Flood cache-based mitigation papers
and adversarial training in SDNs employing active learning techniques.

It is interesting to compare these clustering results 
with the manually defined groups in Figure \ref{fig:taxonomy-goals}, 
in an attempt to discover common traits. 
A major similarity is that 
the ``Detection'' cluster in Figure \ref{fig:dendro_ward} is almost completely
included in the ``Detection Algorithm'' group in Figure \ref{fig:taxonomy-goals}, 
suggesting a clear and well-defined focus on detection methods.
In addition, the two clusters of ``Adversarial Training'' in Figure \ref{fig:dendro_ward} overlap greatly with that in Figure \ref{fig:taxonomy-goals}, 
which is to be expected given the specific algorithms and keywords used in this approach. 
Some other small similarities are visible as well, for specific niches, as in ``time series forecasting/prediction''.

Beyond these similarities, there are a few shared insights. 
The main reason is that the criteria discovered in the automatic clustering process are diverse, 
including specific targeted protocols (HTTP, ICMP, DNS, SYN floods),
generic approaches (Detection, Classification, Clustering), or network environments (e.g., IoT, SDN).
On the contrary, in Figure \ref{fig:taxonomy-goals}, 
there is a single generic criterion considered, namely the technical approach proposed in each article.
It is therefore natural that they lead to fairly different hierarchies. 
This, however, is not detrimental; it only helps the reader to obtain
a more nuanced understanding of the surveyed literature.
 
\section{Training Data: Flows, Graphs and Time series}
\label{sec:data}
\begin{table}
\tiny
\centering
\begin{tabular}{r|llrr}
Dataset & Attacks & Format & Raw Size (GB) & Flows Size (GB) \\
\hline
\href{https://research.unsw.edu.au/projects/bot-iot-dataset}{Bot-IoT}~\cite{Koroniotis-ELSFGCS-2019} & Flood, TCP, HTTP & Raw, Flows & 69.3& 16.7  \\
\href{https://www.caida.org/catalog/datasets/ddos-20070804_dataset/}{CAIDA 2007}~\cite{Alzahrani-SRPIS-2018} & Flood & Raw & 21 & - \\
\href{https://www.unb.ca/cic/datasets/ddos-2019.html}{CIC-DDoS2019}~\cite{Sharafaldin-ICCST-2019} & Flood, TCP, NTP, DNS & Raw, Flows, Time series & 21.34& 0.5  \\
\href{https://www.unb.ca/cic/datasets/ids-2017.html}{CICIDS 2017}~\cite{Sharafaldin-ICISSP-2018} & HTTP, Slow & Raw, Flows &  51.1 & - \\
\href{https://www.ll.mit.edu/r-d/datasets/1998-darpa-intrusion-detection-evaluation-dataset}{Darpa98}~\cite{Lippmann-ELSCOMPNET-2000} & Flood, TCP, Smurf, HTTP & Raw & 5.5 & - \\
\href{https://kdd.ics.uci.edu/databases/kddcup99/kddcup99.html}{KDD99}~\cite{Kayacik-CPST-2005} & Flood, Smurf, PoD & Flows  & - & 0.743 \\
\href{https://www.kaggle.com/datasets/hassan06/nslkdd/data}{NSL-KDD}~\cite{Tavallaee-CISDA-2009} & TCP, Smurf  & Flows & - & 0.02 \\
\href{https://research.unsw.edu.au/projects/unsw-nb15-dataset}{UNSW-NB15}~\cite{Moustafa-MilCIS-2015} & Generic DoS & Raw, Flows & 100 & -  \\
\href{https://research.unsw.edu.au/projects/toniot-datasets}{TON-IoT}~\cite{Alsaedi-ACCESS-2020} & HTTP & Raw, Flows, Time series  & 25 & -\\
TUIDS~\cite{Bhuyan-IJNS-2015} & Flood, Smurf & Raw, Flows & \multicolumn{2}{r}{private}  \\
\hline
\end{tabular}
\caption{A subset of the available DDoS datasets with covered attack types,
         data size and data format.
         The attacks are sorted by attack class and then by attack type
         with volumetric first, protocol, and application last.
         Most datasets provide the raw format and optionally some preprocessed formats such as flows or time series;
         we also included the preprocessed data size where available.
         Dataset names link to the download site.}
\label{tab:databases}
\end{table}

\begin{table}
\centering
\tiny

\begin{tabular}{r | l l l l l l l l}
Precision & NB & RF & k-NN & SVM & (C)NN & LSTM & GAN \\
\hline
Bot-IoT~
  \cite{Koroniotis-ELSFGCS-2019,
  new_Li-ExpertSystemswithApplications-2024,
  new_Malekzadeh-JSupercomput-2025,
  new_Sadhwani-AppliedSciences-2023}
  & 0.9938 & 0.9932  & 1.0000  & 0.9998 & 0.9900  & 0.9999 & 0.9908 \\
CIC-DDOS2019
  ~\cite{Cil-ESA-2021,
  Shurman-IAJIT-2020,
  Amaizu-CN-2021,
  Peng-IEEE-2019}
  & 0.4100 & 0.7700 & 0.8796 & 0.7360 & 0.9952 & 0.9919 &  - \\
CICIDS 2017
  ~\cite{Sharafaldin-ICCST-2019,
    Mahjabin-SpaCCS-2019,
    Peng-IEEE-2019,
    new_Abiramasundari-SciRep-2025,
    new_Alimi-JSAN-2022,
    new_Kim-Electronics-2020}
  & 0.8800 & 0.9800 & 0.9600 & 0.8700 & 0.7087  & 0.9923  & 0.9739 \\
KDD99
  ~\cite{Peng-IEEE-2019,
    new_Kachavimath-ICIMIA-2020,
    new_Xu-ACCESS-2019,
    new_Thangasamy-CSSE-2023}
  & 0.9774 & 0.9890 & 0.9700 & 0.9550 & - & 0.9000 &  - \\
TON-IoT
  ~\cite{Alsaedi-ACCESS-2020,
    new_Sadhwani-AppliedSciences-2023,
    new_Zheng-JNI-2025}
  & 0.6300 & 0.8700 & 0.8500 & 0.3700 & 0.9900 & 0.8300 & 0.9762  \\
\end{tabular}
\caption{Precision on popular DDoS datasets as reported in the literature.
For entries marked with "-" precision was not available.}
\label{tab:databases_results}

\end{table}

We next discuss the various forms DDoS training data can take
before being processed by learning algorithms.
Most IDS analyze the traffic at various network layers.
The first step in data processing covers the capture and storage of network traffic data. This step is essential for reaching high DDoS detection quality.

Some of the most popular public databases are gathered in Table~\ref{tab:databases}.
We can see that, except for KDD,
all data are presented in the raw packet (\texttt{pcap}) format.
Apart from CIC-DDoS2019,
these datasets do not focus on DDoS attacks 
and instead include other types of attacks,
besides normal traffic.
Among DDoS traffic,
flood attacks seem to be the most common,
while TCP and HTTP come in second;
each of these belong to the
volumetric, protocol, and application attack classes, respectively.
We also include, for brevity,
the TUIDS database, which is not currently available to the public,
although it was in the past and is often discussed in literature.
In addition to raw traffic,
some databases also provide preprocessed data
in the form of flows or time series.
We add this information to the table where available,
and proceed to discuss these and other preprocessing formats.

\subsection{Cross benchmark analysis}

We summarize our dataset analysis in Table~\ref{tab:databases_results}
with a brief precision benchmark of AI models against popular datasets.
Precision was chosen as it is a better fit for unbalanced datasets,
overall we have more normal traffic than DDoS.
Precision was also the most common metric used across the literature; where this was not available we used accuracy instead.
We focus on common models, including both
shallow methods such as NB, RF, k-NN and SVM,
and
neural networks such as CNN, LSTM and GAN.
When CNN results are missing, we report plain feedforward Neural Network numbers.
With few exceptions, we can see that the datasets are saturated
and that models tend to overfit. 
Using GANs to enrich the dataset often triggers significant performance drops, as depicted in the last column.

Following the data from Table~\ref{tab:databases_results}
and the experimental sections of the surveyed papers,
we provide some context here
regarding the evaluation practices across different studies
together with the limitations of cross-benchmarking.
First, 
as observed in the table,
precision is not always available as a metric
even though it is the most common one.
Papers are very heterogeneous when it comes to metrics
and any combination of three or four metrics can be found in a given study;
from basic accuracy, precision, recall and F1-score,
to raw metrics such as number of true positive, false positive,
true negative and false negative
sometimes accompanied by sensitivity and specificity,
to more specific metrics such as
fall-out, detection and training time.
Performance metrics are also associated with the different dataset properties,
see Table~\ref{tab:databases},
such as attacks diversity and duration coupled with normal background traffic,
data size and access to raw traffic recordings as required for the preprocessing strategy of each study.
Besides that,
while some studies limit themselves to a pure data science approach
where the datasets are processed by standard or slightly modified AI algorithms
and afterwards a direct metrics comparison is employed,
others make use of these algorithms inside a more elaborate network analysis framework
where, for example, the AI part is only a module in a more complex IDS architecture.

Focusing now solely on algorithmic cross-benchmarking challenges
we note the fact that even if the same method is found across multiple papers,
such as those listed in the columns from Table~\ref{tab:databases_results},
with experiments on the same datasets,
AI training choices such as
different train-test splits,
varied preprocessing strategies due to data structures choices
(see the following subsection) or dimensionality reduction techniques,
and the amount of effort and strategy invested in hyperparameter optimization
can lead to discrepancies in the reported performance metrics.
For neural networks there is also the question of number of layers or activation functions
within the same architecture type.
We thus caution the reader when interpreting Table~\ref{tab:databases_results}
about these issues
and note that we tried to report the results most often
where the experimental section seemed to indicate a thorough and equitable
preprocessing and tuning process.

\subsection{Network traffic structure}

In the network traffic structure, a \textit{packet} is a data unit composed of \textit{header} fields and the proper message (\textit{payload}) to be transferred. Each field from the the \textit{headers} stack, that belongs to a different TCP/IP layer, is necessary for the proper management and routing of the packet toward the destination. Capturing tools such as \texttt{tcpdump} are often used to monitor and store packet-level information of the network.  
When a capturing tool stores and inspects the traffic among multiple machines, the complexity of packet analysis is highly dependent on the number of monitored devices and on the rapidly growing networking speeds. 
Additionally, the detection of real collective attacks (such as DDoS) usually requires a periodic inspection of very large low-level packet sets. For this reason, collection, inspection and management of low-level packet traffic remains a challenge for this format. Therefore, grouping packets into sets that have some shared characteristics becomes a natural idea.


A \textit{flow} represents a packet collection related through a common set of attributes. The attributes include header fields corresponding to transport and application layers (IP addresses, IP protocol, source ports, etc.), as well as various information related to local packet management (e.g.\ input-output physical port). Despite the wide interval of attributes, collection and computation of flow records is an increasingly common capability of usual network devices. The flow records computed on the export device are transferred towards a collector that stores them. Each record has basic flow information and other fields: timestamp for flow start and stop, byte length over the packets in the flow, etc. This way, the flow format will eventually be much more compact than the packet data format.


Regardless of the primary traffic format, three dominant data models allow the diversity of anomaly detection techniques to be applied for detecting network intrusions: tabular, time series, and graph models. 

\noindent Identifying important traffic attributes over each flow (or packets group) is a widely adopted preprocessing step that converts raw flow data into a new \textit{tabular} dataset. After this conversion, many authors proceed to apply particular outlier detection techniques and provide specific interpretations of the results. Some common traffic attributes, retained from the raw traffic, are promoted by many papers to facilitate the prevention of multiple DDoS attacks. Thus, we further exemplify some of the most common features (that we found in the surveyed papers) organized over the class of DDoS attacks they address:
\begin{itemize}
\item Volumetric: source IP,
destination IP, source port, protocol, source packet, destination packet, packet 
    number (identifier), total number of bytes transferred, ToS values, duration of established connections, statistics from counting source MAC and IP address combinations, channel and socket properties.
\item Protocol: source IP,
source and destination IP combinations, IP identifier and fragmentation properties (DF field), TTL values, source and destination TCP port combinations, window size for TCP packets, maximum/minimum packet length, average packet size, maximum/minimum packet forward length.
\item Application: incoming IP addresses, TCP source/destination port numbers, maximum number of sessions, web page access count, number of HTTP request packets per unit time. 
\end{itemize}

By taking into account the timestamp of each flow, a temporal order can be established over the network traffic and the raw data takes the form of a \textit{time series}. With a relevant encoding of the attributes, the DDoS attack detection problem becomes a time series analysis problem. Among the main flow attributes that prove to be worth tracking over time are: new flow record creations per second, average flow duration,
average number of bytes per flow, average number of
packets per flow \cite{hofstede_CNSM_2013}; DNS query rate \cite{Verma-IEEEARES-2016}; statistics based on counting MAC and IP addresses combinations \cite{Mirsky-NDSS-2018}; number of packets, entropy of IP addresses and ports list~\cite{Lent-IEEEA-2024}; number of received packets, received bytes, number of sent packets, sent bytes, port keep alive duration, drop rate for received packets, number of errors for received and transmit packets, etc.~\cite{Das-CSR-2022}.

\begin{figure}
    \centering
    \includegraphics[trim={0.5cm 3.2cm 8cm 0.2cm},clip,width=\linewidth]{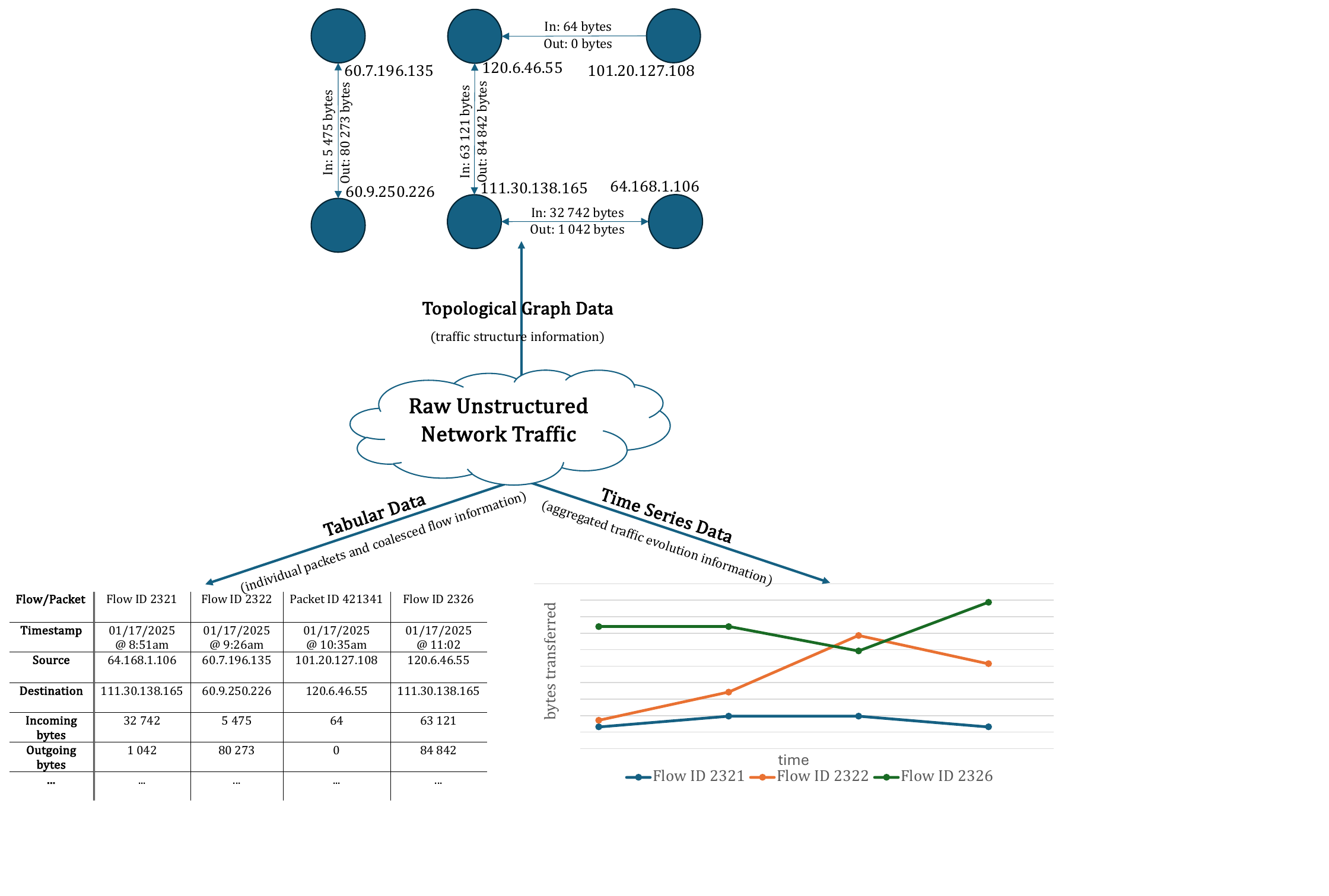}
    \caption{Three representative toy examples of the most popular data extraction/organization techniques used for DDoS applications. From raw unstructured network traffic dumps, we can extract individual packet-level data, or coalesce the data into flows, and we can organize the data into: tabular, time series, and graph/network structured data.
    }
    \label{fig:tab_to_graph}
\end{figure}

The analysis of inherent relations between the flow attributes naturally leads to the third data model based on \textit{graphs}. In general, DDoS attacks are prominently reflected/amplified in particular subgraph patterns where a single node has a high input degree compared with its neighborhood. Even a simple feature table of the flows, that includes a timestamp, source and destination IPs for each flow, can be easily converted into a basic static graph structure where the nodes are the IP addresses and the links are the connections within the network (see Figure \ref{fig:tab_to_graph}).

\section{Detection}
\label{sec:detection}

We next delve into a detailed literature review of state-of-the-art
AI-powered detection methods and solutions against DDoS attacks following the taxonomy described in Figure \ref{fig:taxonomy-goals}. 
Based on their specifics, we group attacks into volumetric, protocol-based, reflection and amplification, and application (Layer 7) attacks. We also summarize the reported detection time of the proposed methods
from the papers that include this information.

\subsection{Volumetric DDoS attacks}\label{sec:volumetric}



Volumetric attacks are connectionless floods in which an attacker uses networks of bots to send a large number of, usually, UDP packets towards an intended victim, disrupting mainly their network infrastructure, but potentially also their network stacks.

In the vast majority of cases, volumetric attacks are treated in a unified manner, just as the other DDoS attacks. This is because well-established datasets, such as CIC-DDoS2019 \cite{Sharafaldin-ICCST-2019}, have recorded traffic from various types of DDoS attacks. As stated in \cite{Cil-ESA-2021}, given the CIC-DDoS2019 dataset, there are two main approaches: (1) detect that a DDoS attack is happening and (2) establish which kind of DDoS attack it is. In general, when discussing synthetic datasets, such as CIC-DDoS2019, the results published in the literature are excellent, exceeding 99\% precision\footnote{All other metrics, such as accuracy, recall or F1-score have similarly high values over 99\%.} for detecting DDoS attacks using only relatively simple AI algorithms. Most probably, these almost perfect results stem from the synthetic nature of the employed datasets, since these were generated in synthetic lab conditions, not real-world situations.

Most research papers obtain similar near-perfect accuracy results. For example, Wei et al.~\cite{Wei_ACCESS_2021} use a mixed architecture composed of an autoencoder and a multi-layered perceptron. In this case, the autoencoder performs dimensionality reduction, while the perceptron performs the actual classification task.

A few-shot model is one that aims to recognize attacks after only seeing very few examples during training. If there are no training examples at all, the model is called zero-shot. This scenario is very useful in the real world because new types of attacks (or variations of the existing ones) appear all the time. Because of that, an important characteristic of an AI model is its ability to detect and respond to new types of attacks. For instance, Shieh et al.~\cite{Shieh-MDPI_AS-2021} propose a human-in-the-loop, semi-supervised solution. New types of attacks are detected in a zero-shot manner. The relevant data samples are sent for validation and manual classification to a human operator. If what is detected is indeed a new type of attack, the information is added to the system, such that, in the future, the AI system will be able to automatically perform the correct classification for this type of attack. The implementation is based on a combination of deep learning, BiLSTM, and GMM techniques.

In most solutions, traffic is modeled using network flow models, making explicit use of the structure of the network. For example, in \cite{Liu-JNCA-2021}, the traffic is modeled through a time-varying network graph model. The regular network traffic is modeled as a fixed set of normalized graphs. An extra element of novelty is that, to measure the divergence from the regular traffic, the authors use a Weisfeiler-Lehman kernel. Compared with previous solutions, the proposed AI system performs better in real-time scenarios and for the classification of various DDoS attacks. In another graph-based approach \cite{Bakar-CN-2024}, a two-stage aggregation technique is proposed, resulting in the construction of graph models for the packet-level and the flow-level traffic. The authors analyze this traffic using CNNs, and obtain state-of-the-art results (especially in terms of the F1-score). The accuracy of the model is proportional to the past traffic window size, a parameter that is important in establishing the topology of the traffic information, but whose increase leads to significantly larger inference times.

Detection of DDoS attacks plays a crucial role in many practical situations. For example, smart camera surveillance systems are critical because disabling them via a DDoS attack allows an on-the-ground operation to proceed undetected. Hence, the network robustness of such systems is essential. The DDoS robustness of surveillance systems is analyzed in the study of Mirsky et al.~\cite{Mirsky-NDSS-2018}, which is based on a set of serially connected autoencoders. Their role is to encode the traffic characteristics and prepare them for anomaly detection methods. For training, the authors use custom datasets encapsulating a set of relevant cyberattacks including recon, man-in-the-middle, and DDoS (SYN flood, SSDP flood, and SSL renegotiation). The proposed solution is highly applicable and focuses on many practical considerations.

\subsection{Protocol DDoS attacks}\label{metode_protocol} 

Protocol-based DDoS attacks generally exploit specific characteristics of network protocols or their software implementation (e.g., TCP, ICMP). In general, the main target of these attacks is the victim's network stack, but the network hardware infrastructure might also be significantly affected.

Among the earliest work in protocol based DDoS detection using ML techniques includes the detection of SIP flood attacks over VoIP networks. First, the work in~\cite{Nassar-RAID-2008} uses an SVM approach and 38 features grouped into four categories to detect telephony flooding and spam attacks. The features are constructed from time-sliced statistics of the signaling messages usually used in SIP for VoIP (especially INVITE messages). Then, the work in~\cite{Akbar-KIS-2014} uses ML techniques such as NB and DT on a dataset composed of both spatial and temporal engineered features extracted from raw network packets. In this setting, temporal features are variations in the number of INVITE, ACK, and BYE messages captured in traffic while spatial features are computed by the entropy of caller IDs in a time window. The results reported in both papers are above 90\% in terms of detection accuracy.

As mentioned in previous sections, many solutions in the scientific literature
jointly handle many DDoS attacks,
developing general detection and protection methods working with multiple attacks.
This is also reflected in the databases used for training and validation \cite{Sharafaldin-ICISSP-2018,Sharafaldin-ICCST-2019}.
More precisely, even if individual attacks differ, technically depending on the
specifics of the protocol exploited, the types of extracted features and the used detection methods are usually generic,
and can be applied to a wide range of attacks.
For example, in \cite{Zhao-CS-2024}, the datasets used for evaluation
contain 89 different variations of DDoS attacks.
Even though some works explicitly refer only to certain types of attacks,
usually TCP or TCP-SYN floods \cite{Dimolianis-ICIN-2021,Guo-SCN-2022,Das-CSR-2022},
this is largely due to the test environment and the datasets used for evaluation,
and not to the methods themselves being specialized for certain types of attacks. In contrast, specializing on certain types of attacks can be inherently problematic,
since several types of attacks are not represented well enough in the current databases
to be properly modeled and analyzed \cite{Das-CSR-2022}.

Most of the works analyzed in this section start from
features extracted from flows, such as the number of packets,
number of bytes, and duration \cite{Khashab-Netsoft-2021},
together with identification elements
(source address, source port, destination address, destination port, and protocol).
The values are taken at periodic intervals to monitor their evolution in time.
Several works mention the particular relevance of these differential parameters: Khashab et al.~\cite{Khashab-Netsoft-2021} use the number of flows to the same host and the same port, respectively, from the last 5 seconds;
Das et al.~\cite{Das-CSR-2022} take the differences between the number of packets and bytes at successive intervals of 5 seconds;
Zhao et al.~\cite{Zhao-CS-2024} use a set of 5 sliding time windows, ranging from $0.1$ to $10$ seconds.
The papers above use manually defined features, but other approaches use automatic methods for extracting relevant features,
such as those based on autoencoders \cite{Ko-JISA-2020},
feed-forward networks \cite{Liang-NDSSS-2021}
or attention mechanisms \cite{Guo-SCN-2022}.

A special case is the work of Das et al.~\cite{Das-CSR-2022}, which uses only parameters collected
from the ports in the network infrastructure,
without considering features extracted from flows.
This is considered an advantage, as flow-level statistics are dependent on the network topology,
and the associated traffic, which could impact the generalization of the results \cite{Das-CSR-2022}.
In this case, once an attack is detected, the origin of the attack is identified through a localization process,
which analyzes traffic on all ports and identifies the switches most affected,
considered to be the closest to the source of the attack.

The methods used for detection are varied, but most of them are based on supervised learning algorithms:
SVM, RF, DT,
neural networks with MLP or CNN architectures \cite{Das-CSR-2022,Daoud-DKE-2023,Khashab-Netsoft-2021,Zhao-CS-2024}. Along the same lines, the work in~\cite{Tripathi-CVHT-2018} uses a simple OC-SVM classifier to detect particular types of DHCP starvation attacks. The approach is rather simple and uses four features extracted from DHCP traffic: the number of DISCOVER, REQUEST, and DECLINE DHCP messages and the total number of ARP messages in a fixed time quanta. These statistics are enough to reach near-perfect detection accuracy.
The majority of works use and compare several types of classifiers,
with sometimes extremely different results, without a consensus on the best method
(for example, Das et al.~\cite{Das-CSR-2022} report an accuracy of 51\% with MLP networks).
An exception to this is the articles in which the classification method is linked
with the feature extraction method, such as \cite{Ko-JISA-2020},
which uses autoencoders for feature extraction simultaneously with classification.
Most classification methods are traditional feature-based approaches, in different flavors. An exception is the approach of Guo et al.~\cite{Guo-SCN-2022}, which uses bidirectional GRU models,
coupled with a hierarchical attention mechanism.

Many of the works that use supervised methods report
detection accuracy values of over 99\%, often over 99.9\%,
for example \cite{Dimolianis-ICIN-2021,Zhao-CS-2024,Das-CSR-2022,DoriguzziCorin-IEEETNSM-2020,Guo-SCN-2022}.
It is not clear to what extent these performance levels transfer to real test environments,
or are just a consequence of using the limited set of publicly available databases,
or of using certain software tools for simulating attacks in tests.
However, we note that many works mention evaluating the results on real, private datasets
recorded from ISPs \cite{Zhao-CS-2024, Dimolianis-ICIN-2021},
which increases the level of confidence in the proposed solutions.

Besides attack detection, a large number of works also consider
the other stages required in the entire mitigation pipeline, proposing
various optimizations and improvements.
A popular topic is optimizing the number of rules required for traffic filtering,
which may increase proportionally with the number of addresses involved in the attack,
potentially becoming very costly \cite{Dimolianis-ICIN-2021,Zhao-CS-2024}.
Several approaches to consolidate the filtering rules are proposed,
which rely on the similarity of some attack patterns to counter them with as few rules as possible.
In \cite{Dimolianis-ICIN-2021}, a method based on genetic algorithms is proposed,
which can reduce the number of rules and signatures used in detection by up to 99.99\%, according to the authors. We note, however, that this work considers only SYN Flood attacks.
These consolidated rules are implemented with programmable XDP enabled firewalls;
since optimizing the rules is a slower process, the SYN cookies technique is used
as a temporary protection measure for quick response, until the rules are updated.

The concept of ``DDoS attack family'' is proposed in \cite{Zhao-CS-2024},
aggregating attacks that allow similar defense strategies.
This is done by evaluating the differences between signatures of different attacks,
representing the relationships between them as graphs,
followed by partitioning them into sub-communities with specific algorithms.
The approach also allows for adaptation to new types of attacks, not encountered in the design phase.
The analysis of the authors shows that 89 types of attacks can be grouped into just four families, namely: i) attacks to network Layer 4 or higher; ii) TCP attacks without a response from the victim; iii) TCP attacks with a response from the victim, and; iv) connectionless attacks.
As proof of the efficiency of this solution,
to counter all attacks in category iii), only two rules are needed,
which are based on counting RST packets and packets having both fields SYN and ACK set.

The possibility of adapting to new types of attacks is also addressed in the work of Doriguzzi-Corin et al.~\cite{DoriguzziCorin-CS-2024},
who investigate a federated learning approach to incorporate data and models from multiple clients, while avoiding data centralization.
In \cite{Liang-NDSSS-2021}, a semi-supervised learning method is introduced,
which is appropriate in situations where attacks are poorly labeled or even unknown.
The method is based on clustering the data and then projecting it into a reduced space,
aiming to preserve the cluster membership for data with available labels.

As mentioned above,
even though most methods do not specialize on a specific attack,
some do specialize.
In the following, we address these specialized methods.
For
TCP Connection/SYN Flood,
in addition to the articles presented in the previous section,
those using the CIC-DDoS2019 dataset \cite{Sharafaldin-ICCST-2019} implicitly address SYN Flood.
Thus, the works of Salahuddin et al.~\cite{Salahuddin-TNSM-2021}, Wei et al.~\cite{Wei_ACCESS_2021}, and Shieh et al.~\cite{Shieh-MDPI_AS-2021}, described in Section \ref{sec:volumetric}, are relevant to the subject.
PoD attacks are characterized by packets having, after reassembly, a size greater than the maximum size of 65535 allowed by the protocol.
This attack is explicitly addressed in \cite{Abdollahi-WPC-2020}, which detects it using
a threshold for the number of packets in a time window with a size greater than a certain limit.
Next,
the BGP
attack exploits the vulnerabilities of the BGP protocol, and is based on sending malicious routing updates.
In \cite{McGlynn-INFOCOMWKSHPS-2019}, these attacks are detected using two autoencoders to extract the essential features from the AS-specific information.
Finally,
Smurf DDoS attacks involve spoofing the source IP addresses of ICMP packets sent to broadcast addresses,
which causes a large number of responses to be sent to the devices holding the spoofed IP addresses.
Nowadays, this type of attack is simply avoided by disabling the broadcasting option on network devices.
For this reason, there are few works addressing this attack in the literature,
and most of them use relatively simple approaches,
such as \cite{Bouyeddou-ICCTA-2018}, which harnesses the Kullback-Leibler divergence to identify anomalies.
Instead,
when it comes to the ICMP protocol,
interest in Ping Floods have resurfaced recently in the context of IoT networks~\cite{Almorabea-IEEEA-2023}.

\subsection{Reflection and amplification DDoS attacks}

RA-DDoS attacks enhance the previously described volumetric and protocol DDoS attacks.
Historically, reflection and amplification attacks represented a subset of the broad class of volumetric attacks and are characterized by an attacker who can selectively use a relatively small amount of traffic employing bots, vulnerable protocols and services to generate a considerable amount of traffic toward an intended victim. These types of attacks usually target services such as DNS and NTP, as well as both standard transport protocols: mostly UDP, but also TCP. An attacker could just reflect traffic toward a victim, but by far, the most dangerous attacks are those where the attack volume is amplified manyfold. Although RA-DDoS attacks were initially based exclusively on UDP, subsequent advances~\cite{Kuhrer-USENIXWOOT-2014, Bock-USENIX-2021} showed how TCP can also be used to reflect and amplify malicious traffic. Due to these characteristics, reflection and amplification attacks are sometimes treated separately in the DDoS volumetric attack literature \cite{Mirkovic-ACMSCCR-2004}. 

Most classic techniques used to detect and mitigate RA-DDoS attacks rely on changing the network's configuration parameters or the targeted services. This mitigation technique bears the name System and Network Hardening \cite{Rossow-NDSS-2014, Kuhrer-USENIX-2014, Kuhrer-USENIXWOOT-2014, Bock-USENIX-2021}. Anticipating AI-based techniques, a particular line of work \cite{Verma-IEEEARES-2016, Abdul-IFS-2020, Wagner-ACMCCCS-2021} follows the idea of collecting simple statistics about network traffic to detect and then suggest mechanisms to stop RA-DDoS attacks. 


Among the first papers to use AI techniques to detect RA-DDoS attacks, we mention here works \cite{Meitei-ICIA-2016, Chen-IEEEDSC-2017} that use several machine learning techniques such as DT, MLP, NB, and SVM to detect DNS RA-DDoS attacks.

In the following decade, Anley et al.~\cite{Anley-CS-2024} turned to modern AI techniques, such as deep CNNs together with transfer learning, to enhance the generalization capabilities of AI models by combining data from multiple freely available DDoS datasets. The recent literature views the problem of generating realistic RA-DDoS attacks as a central one in the development of reliable detection methods. For this reason, the work of Mathews et al.~\cite{Mathews-CMSIE-2022} proposes a robust RA-DDoS method that uses neural networks and adversarial learning techniques, such as EAD and TextAttack, to generate counterexamples for the AI model, which will not be correctly labeled as attacks. Such enhancements lead to better generalization in practical datasets. In the same spirit, a method that uses a GAN-based discriminator for anomaly detection is proposed in \cite{Lent-IEEEA-2024}.
This uses GRU-type neurons, analyzing traffic as a time series.
Features such as the number of transmitted packets, the number of bits and
entropy of IP addresses and ports are extracted for each one-second time window,
and are aggregated into overlapping 10-second time series for analysis. The generator network learns to synthesize normal traffic flows with features very similar to those from the training set, while the discriminator improves the anomaly detection capability. In addition to the detection module, the authors also design a mitigation module that identifies a victim as the IP address that receives the largest number of flows in a time window detected as an anomaly, and subsequently blocks the IP addresses that send traffic towards it.

Regarding the features used in the detection process, several approaches are possible.
One of them involves selecting a subset of features through different methods,
as in \cite{Shurman-IAJIT-2020, Wei-arXiv-2023}.
Another option, found in \cite{Lyu-TNSM-2021, Lent-IEEEA-2024}, is to build new features from a time window
containing a variable number of packets or flows.
The former option allows for inference as soon as the information about a flow is stored in the database,
whereas the latter requires building new features only after all the flows corresponding to a time window are available and stored. Amaizu et al. \cite{Amaizu-CN-2021} use the Pearson correlation coefficient for feature selection and then jointly train two neural networks with different architectures to predict the type of attack for the flows in the CIC-DDoS2019 dataset. The attack classes are not limited to RA-DDoS types, as they also include general volumetric attacks.






As with other types of attacks, RA-DDoS attacks are usually addressed collectively in the scientific literature.
A notable exception is the work of Lyu et al.~\cite{Lyu-TNSM-2021}, who develop detection methods for RA-DDoS attacks based on the DNS protocol.
They propose a hierarchical graph structure with a dynamic retention policy to model traffic at three different levels:
host, subnet, and AS.
Then, two different methodologies are used to detect traffic anomalies at each level.
The first method uses static detection rules for each scenario, while the second uses IF and OC-SVM models to detect anomalies
based on the following time-windowed features: variance of packet size, number of internal hosts queried by each external entity,
average number of packets sent in queries to each internal host, and variance of the number of packets sent in queries to them. The authors further used two of these features to separate anomalous hosts in scanners and flooders.

Because of the time-dependent nature of monitoring network traffic quantities, recurrent neural structures have been used with great success for RA-DDoS detection. Shurman et al. \cite{Shurman-IAJIT-2020} proposed two approaches to detect RA-DDoS attacks.
The first one is a simple framework consisting of a database with known attack signatures and an anomaly-based detector. Here, the signature database is updated when a packet whose characteristics deviate from the normal ones is analyzed by the anomaly-based detector.
The second approach uses LSTM-based networks with varying number of LSTM layers to detect RA-DDoS attacks from the CIC-DDoS2019 dataset.
An RF classifier is used together with the GINI impurity to select a subset of features for training the LSTM-based model. Another LSTM-based architecture is proposed in \cite{Wei-arXiv-2023} to detect RA-DDoS attacks as time series anomalies within a time window containing a configurable number of data flows.
The training process uses a subset of the features from legitimate data flows, and then a threshold on the reconstruction errors obtained by the LSTM autoencoder is derived.
The proposed architecture contains an encoder, in which the output of each LSTM cell component is ignored, followed by a decoder,
where the outputs of the component cells represent the reconstructions of the input data. The authors reported separate results for RA-DDoS attacks based on DNS, LDAP and SNMP.
A similar LSTM autoencoder is used in \cite{Said-ACMSQSWMN-2020},
but here, latent representations are used to train an OC-SVM model, because the simple threshold-based approach did not achieve satisfactory performance.
In both papers, only instances labeled as normal are used in the training process.
While the work of Wei et al.~\cite{Wei-arXiv-2023} was used to detect three types of RA-DDoS attacks, Said et al.~\cite{Said-ACMSQSWMN-2020} presented a model trained to detect several types of attacks, including DDoS.

A recent survey \cite{Ismail-CS-2021} focuses particularly on RA-DDoS attacks, and yet another one \cite{Wabi-CS-2024} showcases the recent work in studying RA-DDoS attacks in the context of SDNs.

\subsection{Application-level DDoS attacks}

We provide an overview of the current state of research as well as some possible future research directions concerning application-level DDoS attacks. Alongside the information present in the surveys section and the works concerned with Layer 7 Application attacks, we can distinguish three broad types of attacks, each of them having multiple subtypes.
More precisely, we discuss detecting HTTP attacks, especially flooding attacks, followed by detecting slow HTTP attacks (Slowloris and R.U.D.Y.). Lastly, we discuss hierarchical DNS server attacks.

As compared with previous DDoS attacks, application layer DDoS usually targets the operating system, software stack and hardware compute resources of the target victim machines. The network infrastructure is under some strain, but it is usually not the main target of the attack. In many application layer DDoS attacks, the issue is not necessarily the large number of packets transferred,
but the compute burden due to the malicious content of the messages and how these are interpreted and processed by the victim's software.

\subsubsection{HTTP/2, HTTP/HTTPS flood}

HTTP flood attacks refer to the scenario where an attacker uses malicious HTTP requests to target an HTTP server with the goal of making it unresponsive to legitimate traffic.

Because the HTTP requests involved in this class of attacks are perfectly normal requests, those attacks are difficult to tell apart from the legitimate user traffic. Wang et al. \cite{Wang-SCIS-2014} propose a method to classify HTTP flood attacks into two types: random and perfect knowledge. In perfect knowledge ones, the attackers know the structure of the website and attempt to mimic the behavior of a legitimate user. For the random ones, this assumption does not apply. This difference is important for the proposed solution, because it uses the learned distribution of the requests to identify the legit users. The solution, named HTTP-SoLDiER, is based on the correspondence between the individual clicks generated by users with the general distribution of the users of the site. Based on deviation theory, the solution can differentiate legitimate users from attackers by computing the probability of navigation, with respect to general interest and popularity. The critical point of the paper is the fact that, because the content of the pages and the structure of the site can change, the probability must be dynamically adjusted. To prevent the corruption of statistics during an attack, the authors propose a correction algorithm. This algorithm, called EWMA, estimates the popularity of pages in real-time, based on current state and historical data, thereby reducing the bias introduced by the attack itself. A noteworthy observation, also mentioned in previous literature, is that the general distribution of page popularity follows a Zipf law~\cite{Piantadosi-PBR-2014}. Because the complexity of their proposed solution is computationally $O(M)$, where $M$ is the number of pages, the computation requirements are not significant. Alongside the correction algorithm, the success rate in detecting malicious traffic is 99\% for random HTTP floods and 77. 9\% for perfect-knowledge attacks. The downside of this solution is that some websites only have a single page/URL and the so-called circular perfect-knowledge attacks, where the attackers mimic the behavior of a legitimate user by rotating several pages, are significantly harder to detect.

Anomaly detection can be used to detect HTTP flood attacks. For example, Najafabadi et al.~\cite{Najafabadi-FLAIRS-2017} propose an anomaly detection mechanism, based on text mining, to detect HTTP flood attacks. To be close to a real-world scenario, the benign data is generated by collecting the internal traffic in the university network and the attack traffic is generated during a penetration testing session designed to simulate a real attack. Despite that, the dataset is still a synthetic one, generated using software tools.
The proposed mechanism works as follows: a document is defined as a request-response series corresponding to an HTTP GET. The requested resources are regarded as words/tokens. Features are extracted from each document by using bi-grams. OC-SVM is trained using benign data, to model the normal behavior of users. Then, this model is tested on attack data to measure performance. 

Inspired by bioinformatics, some algorithms for the detection of HTTP floods are tested by Sreeram et al.~\cite{Sreeram-ELSACI-2019} and Sree et al.~\cite{Sree-NCA-2019}. The study of Sreeram et al.~\cite{Sreeram-ELSACI-2019} is a theoretical one, using the CAIDA dataset. The employed metrics include session time, the maximum number of sessions, the minimal time between requests, and the number of different packet types observed during this session window. The metaphor-based metaheuristics Bat algorithm is used for detection in both papers. The algorithm simulates bats that fly in a search-space, and the distance and direction of the search are dynamically corrected based on the intensity and frequency of the echolocation pulses.

Sree et al.~\cite{Sree-NCA-2019} propose using the same Bat algorithm as a clustering method to find sources of HTTP flood attacks. This paper focuses mainly on the cloud computing aspects, detailing the practical side of the implementation. The method used in the paper, fuzzy Bat clustering, combines FCM and the meta-heuristic algorithm inspired by the Bat algorithm. This combination is designed to tackle the problems of FCM, such as the selection of the cluster center, detecting unknown attacks, and avoiding local optima. The proposed software solution consists of four components: log collection, data storage, processing, and the detection method itself. Data are collected from virtual machine logs, network logs, and access logs. Once stored, the data is cleaned by pre-processing, and fed to the detection module. The parameters used for classification and identification are: the number of requests in a time window, the frequency of the requests, response times, request similarity, and the number of times the same request is repeated. The testing framework is private, in the cloud, using OpenStack. The malicious traffic is generated using software tools.

DDoS HTTP mitigation methods are grouped by Park et al. \cite{Park-IEICETIS-2021} into two categories: destination level mitigation (implemented by the HTTP server) and network level mitigation. The work proposes an architecture designed to mitigate both DDoS HTTP flood attacks. The solution uses a combination of SDN and rules implemented on the web server. This way, by using both mechanisms, after initially identifying an attacker at the server level, the subsequent requests can be interrupted promptly, before reaching the web server, reducing the impact of the attack. The paper presents the general architecture without providing details about the implementation, the detection algorithm, or about the restrictions imposed on the infrastructure during the mitigation process. The hardware solution consists of two OpenFlow switches connected to each other, one placed between the attacker and the server and the other one close to the server. The switches are connected to an SDN controller. The controller is named HDFD and it has multiple roles in mitigating the attack damage. The protection flow operates like this: once a request is classified as suspicious by the server-level detection component, this connection is sent to the SDN. The switches interrupt the server connection and the SDN mimics the server, responding using headers similar to those of the server, to trick the attacker. Those headers are useful to restore the connection if the user is in fact legitimate. If the user does not manage to pass the test that was set up by the SDN, then the rules of the switches are updated, blocking the suspicious connection. 

\subsubsection{Low \& slow attacks: R.U.D.Y.~and Slowloris}
In HTTP flood attacks, the sheer number and size of requests are used to overload the server. R.U.D.Y.~and Slowloris attacks are more sophisticated because they use specially crafted connections. More precisely, Slowloris transmits the data very slowly, and R.U.D.Y. fails to complete the post requests. The similarity is that they both mimic clients with slow Internet connections, which can clog the computational resources of the victim's web server. 

The performance of a few basic ML algorithms in detecting R.U.D.Y.~attacks is compared by Najafabadi et al.~\cite{Najafabadi-FLAIRS-2016}. This paper offers a scalable solution and compares three different detection algorithms: k-NN, C4.4D and C4.5N DTs. The attribute selection is decided by comparing 10 different methods, such as: F-metrics, geometric means, mutual information, Fisher scores, Kolmogorov-Smirnov statistics, Chi Squared, S2N, among others. Tests are performed on traffic taken from SANTA, a dataset composed of commercial traffic collected in the network of an ISP. It is worth mentioning that the collected traffic is bidirectional. By using Netflows, the author can group the data into sessions corresponding to complete round-trips. The selected parameters are divided into three categories: session similarity, periodicity, and the speed corresponding to requests and responses. To compare the predictive models, the author uses ANOVA analysis, thereby measuring the performance with the full feature set versus the characteristics selected in the comparison process. 

Another method to detect the low \& slow DDoS HTTP attacks is by using time series, as shown by Vitalii et al. \cite{Vitalii-IJETER-2020}. The idea of this work is to use time series to predict normal user behavior. Using these statistical predictions, a projected trajectory of the future actions performed by the user is computed. Those computations are based on a set of selected parameters. This mechanism is often found in literature, based on four components: a module for collecting traffic, data storage, pre-processing of data and traffic parameter computation and, finally, a detection module. The authors emphasize the importance of detecting attacks early. They also highlight the impossibility of completely preventing attacks, due to the fact that they mimic normal user behavior to a large extent. It is worth mentioning that, because slow attacks are the opposite of flood attacks, the parameters used for computations are different from the ones used in the previous section. Predicting the latency of the network packets enables the detection of slow DDoS attacks, based on an algorithm that identifies the unknown future values in a traffic parameter time series. The method mixes self-learning and statistical analysis, based on the existence of a sufficient volume of statistical data about the attacks. 

A solid state-of-the-art analysis is given in Rios et al. \cite{Rios-ACMSIGAPP-2024}. The work also proposes a new way of detecting slow HTTP DDoS attacks, especially Slowloris. 
Because Slowloris traces are publicly unavailable, the authors collected data in three different media: emulated, LAN, and MAN. To facilitate the 10-fold cross-validation, the authors combine four data traces (emulated, LAN, MAN-IF and MAN-Pal) in a single dataset. This dataset is randomly split 75\%/25\% for each of the 10 folds. The original solution is a combination of three methods: FL, RF and ED. Because of this, the proposed method is called FRE. FL is initially used for classification, RF for a detailed analysis, and ED is a fail-safe mechanism to break the ties. If the first two methods share the same result, it is considered final, otherwise ED is used to compute the minimal distance between features, comparing the normal and attack patterns. Nine ML algorithms are analyzed in the paper using two scenarios, one with generic hyper-parameters and the other with optimized hyper-parameters. The analyzed algorithms are:
GNB,
MNB,
XGB,
LGBM,
k-NN,
MLP,
SVM,
DT,
and
RF.

\subsubsection{DNS server query attacks}

This type of DDoS attack targets the DNS server infrastructure of a network. It can target root servers, top-level-domain servers, name-servers, and sometimes even resolvers.
Such attacks are hard to detect and stop because, often, these servers are not under the control of the administrators and developers of the services that are the ultimate targets of the attack. 

There is a common frustration caused by the difficulty of any major upgrade of the DNS protocol and the slow, difficult or missing coordination among the DNS providers to implement the proposed DNS infrastructure and solutions~\cite{Wang-KSII-2012}. This is why, in time, public institutions and private enterprises have migrated towards implementing private DNS servers to facilitate testing and data gathering, and to benefit from the flexibility of a custom implementation. Most solutions recognize the difficulty in changing or updating the DNS protocol and focus instead on backward-compatible optimizations that can be implemented on a subset of servers to diminish the currently known issues~\cite{Mahjabin-SpaCCS-2019}.

As the technology advanced and the adoption of AI increased, the direction of research in DNS flood prevention also changed. It is worth mentioning, though, that the works proposing AI-based solutions start from the same premise, more precisely, the control over the DNS protocol, either in the resolver part or at another point in the chain. 


The work of Lyu et al. \cite{Lyu-TNSM-2021}, also covered in the RA-DDoS section, proposes a new method to detect DDoS DNS attacks by implementing a hierarchical graph structure. The solution can classify, with a high degree of certainty, the following attacks: scans, DNS floods, and slow attacks. This paper presumes control over the DNS resolvers and of the top level domains. The solution is implemented on the network boundary, by duplicating the traffic using an SDN switch. The traffic is subsequently parsed and only the DNS-related one is kept. A pre-processing module is implemented using the DPDK and Intel NFF-Go libraries. The parameter values are combined and estimated to cover the whole range of attacks. The new data is added to a dynamic graph to be analyzed. The resulting graph has a hierarchical structure that is able to establish a link between the attack targets and the IP addresses, IP classes, and the zones where the attacks originate. Hence, the anomaly detection decisions can be stratified, leading to better overall precision.

\subsection{Detection delays}
\begin{figure}
    \centering
    \includegraphics[width=\linewidth]{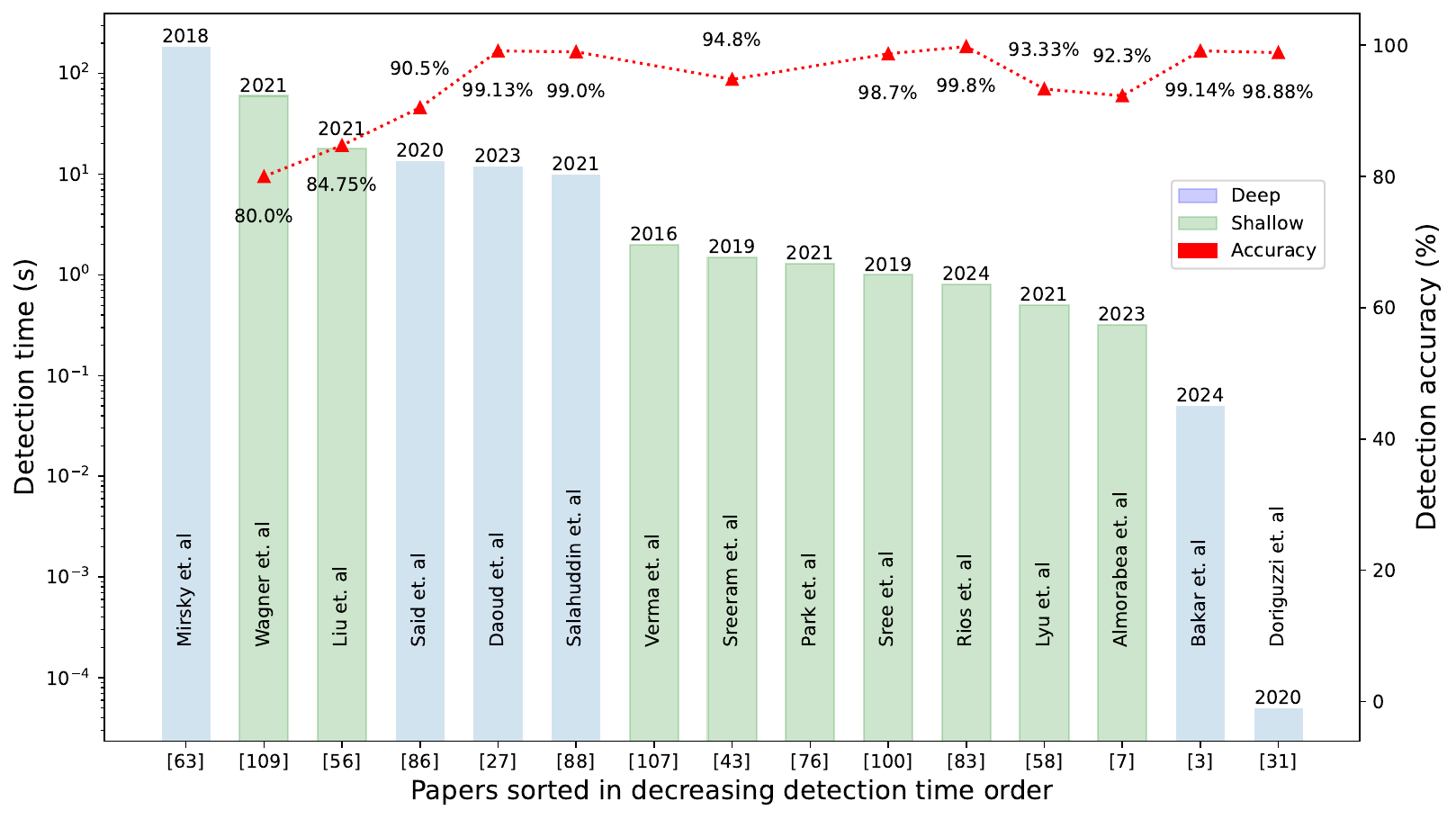}
    \caption{Papers that reported DDoS detection time (left) and
    their associated accuracy score on testing data where available (right).
    The papers are sorted in descending order
    with citation indexes on the bottom axis,
    author names on the bar (including DOI links)
    and publication year at the top.
    The bars are color coded: blue for deep learning and green for shallow learning methods.
    The red line shows top reported accuracy results.
    }
    \label{fig:inference-time}
\end{figure}

The damage caused by a DDoS attack is highly determined by the necessary duration of a given system to detect and mitigate the attack. Thus, in addition to accuracy or other learning quality indicators, a particularly important feature of detection systems is the detection time (or speed) of an incoming attack. Even though this aspect is quite insightful, we found a relatively small amount of papers that rigorously report the detection time or speed of their particular methods or systems. Referring to detection time, most authors report some values for their particular context (dataset dimension, hardware configuration, methodologies, etc.) that are measured in seconds, and in rare cases, in-flows per second or samples per second.

Few papers report the delay between mounting the DDoS attack and 
its detection.
In Figure \ref{fig:inference-time}, we plot a comparison of the papers that do.
We tried to keep a fair comparison
where we take the best results presented by the authors
when performing detection on testing data, including the raw packet processing time.
Some authors do not discuss or include this last bit of information.
Where available, we also added the associated accuracy for the reported detection time.
From the publication years at the top, we can observe that recent years show
an improvement in detection times, even though this is not a strong trend.
It is also visible that shallow learning methods dominate.
Regarding the accuracy-time trade-off, we see very high accuracy values,
which can lead to simple selection criteria:
choose the method with the fastest detection time.
It will be interesting to see how the accuracy quality fares
when using the trained model across different databases,
which we discuss as a future direction at the end of this survey.

\hlight{R3-C4}{
A final point on the accuracy values from Figure \ref{fig:inference-time} is that they are very high in general, with several results approaching near-perfect performance. Rather than taking this as definitive evidence that all available datasets are overfit, we interpret it more cautiously as a sign that several widely reused public benchmarks may not fully reflect the difficulty of real-world deployment conditions, where effective DDoS detection is considered an open difficult problem. This observation is consistent with our broader concern regarding dataset realism and cross-dataset generalization, as also discussed in Sections \ref{sec:volumetric} and \ref{metode_protocol}.}

\subsection{Existing DDoS detection solutions}

We notice that a wide range of AI-based, traditional rule-based, and hybrid model-based software and hardware solutions exist that can detect and, in some cases, mitigate DDoS attacks. We give a brief description of both commercial and open-source solutions, next.

In the case of commercial solutions hybrid models predominate, details of the utilized AI detection methods are often opaque, hidden below several layers of data processing, and therefore we cannot report in this survey details of the implementation. Netscout Arbor\footnote{\url{https://www.netscout.com/resources/solution-briefs/why-netscout-arbor-ddos-attack-protection-solution-is-better}} is a complex, multilayered commercial product that ensures enhanced protection against DDoS attacks, offering cloud and on-premise implementations. It was designed to mitigate large scale volumetric attacks, filtering malicious traffic before it affects targeted devices. The on-premise platform has the capability to collaborate with the cloud-based one when large DDoS attacks overwhelm it. Cloudflare\footnote{\url{https://www.cloudflare.com/ddos/}}, another popular solution, integrates real-time website, application and network protection and mitigation against DDoS attacks, guaranteeing scalability. AWS Shield\footnote{\url{https://aws.amazon.com/shield/}}, a service designed to protect applications that run in AWS cloud, offers the advantage of a seamless, full integration with the AWS services. From the category  of commercial products, we also mention Imperva DDoS Protection\footnote{\url{https://www.imperva.com/products/ddos-protection-services/}} and Radware DefensePro\footnote{\url{https://www.radware.com/products/defensepro/}}, which protect against layer 3, 4 and 7 DDoS attacks.

We also notice a few traditional rule-based open-source solutions that are used for DDoS detection and mitigation. Gatekeeper\footnote{\url{https://github.com/AltraMayor/gatekeeper/wiki}} offers a scalable design with a geographically distributed architecture and a centralized network traffic management policy, while FastNetMon\footnote{\url{https://github.com/pavel-odintsov/fastnetmon?tab=readme-ov-file}} can analyze protocols like NetFlow, sFlow, and SPAN.

Some of these platforms report using machine learning models in order to detect increasingly complex DDoS attacks, but, to our knowledge, none of these offer details about the models used or their implementation.

\section{AI Generated DDoS Traffic}\label{sec:ai-traffic}

Training robust systems for DDoS attack detection involves the use of a rich dataset of examples with a high degree of variety. Since obtaining representative data is not always easy, a few methods in recent literature 
have resorted to the use of generative models to obtain synthetic training data, with the objective of replacing the tedious process of collecting real data. Among the employed models are GANs and VAEs.

Some methods have employed autoencoders as the main tool for improving network IDS by handling noisy or incomplete data. For example, Hashemi et al.~\cite{Hashemi-IEEE-2020} propose a method to improve DDoS detection systems using autoencoder-based architectures to eliminate the noise (denoising autoencoders). The authors studied two architectures, RePo (Reconstruction from Partial Observation) and RePo+. The first one aims to reduce the adversarial traffic reconstruction error by training the model for traffic reconstruction from incomplete data, which leads to a better classification between benign and malicious attacks. RePo+ uses stochastic inference with multiple random masks to reduce the chances of adversarial examples fooling the system. The result of using denoising autoencoders for improving network IDS is the increase of detection accuracy by up to 45\% in the adversarial context. Overall, the detection of attacks improved by 29\% on the packet level and 10\% on the stream level, compared with the results obtained by methods such as Kitsune, DAGMM, BiGAN. 
In a more recent yet similar approach, Saka et al.~\cite{Saka-TC-2023} extend the use of autoencoder architectures and create entirely new traffic samples with the TVAE, specifically designed for generating synthetic tabular data. As opposed to standard VAEs, which struggle with handling mixed data types, TVAE learns a probabilistic latent space representation of the data, ensuring that the newly generated samples maintain the original dataset's structure and dependencies. Despite the success in mirroring the real data distribution, the synthetic dataset used to train a RF Classifier registered a performance drop in accuracy to 93\%, compared with the 98\%-99\% achieved by GAN-based models.

\begin{figure}
\centering
    \includegraphics[width=0.7\textwidth, angle=0]{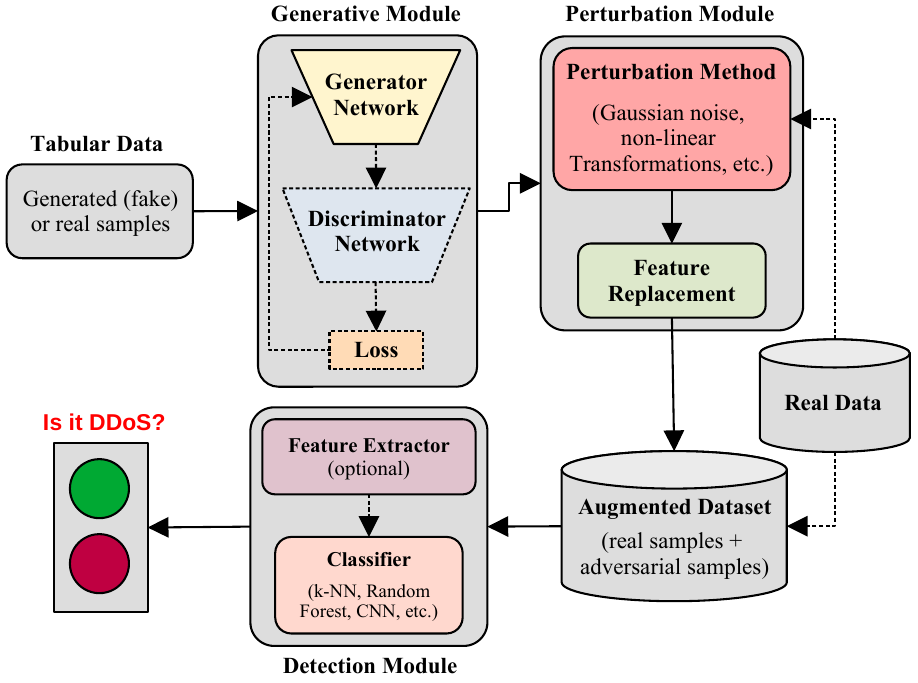}
    \caption{A generic GAN-based architecture for a network intrusion detection system. The input typically consists of tabular data, which can be either real samples from the dataset or generated fake samples. This data is sent through the Generative Module, which comprises a GAN designed to create realistic synthetic data that mimics benign traffic or DDoS attacks. This module can also contain preprocessing steps, such as ternary encoding~\cite{Aleroud-CIW-2020} or conditional information~\cite{Saka-TC-2023}. The role of the Feature Perturbation Module is to alter the previously generated samples (creating malicious examples that look benign or slightly perturbed versions of the original attacks)  using Gaussian noise, non-linear transformations, or other methods. The output of this process is an augmented dataset, which combines the original real samples and the newly perturbed adversarial samples. The augmented dataset is further used to train and/or evaluate a detection system. In the training scenario, a classifier is used to measure the robustness against the adversarial examples.}
    \label{fig:gan-based-nids}
\end{figure}

While autoencoder-based methods focus on improving the quality of generated data, GANs offer an alternative approach by generating realistic data which enhances model robustness. 
In Figure \ref{fig:gan-based-nids}, we illustrate a typical GAN-based approach for improving the efficiency of intrusion detection models.
Dual approaches, such as the ones in~\cite{Saka-TC-2023}, bridge the methods by also highlighting the strengths of GAN variants, specifically CTGAN and CopulaGAN. CTGAN uses conditional input during training to model feature dependencies, while CopulaGAN employs preprocessing with Gaussian copula transformations. 

An earlier application of GANs, SDN-GAN~\cite{Aleroud-CIW-2020}, represents a specialized method designed to generate synthetic attacks targeting software-defined networks. Unlike traditional GANs, this variant employs a binary encoding for representing features. Moreover, it goes beyond the classic GAN framework (based on a generator and a discriminator) by incorporating a new component, the intrusion detector, which creates a feedback loop that produces adversarial samples that can bypass detection. Without GAN-generated adversarial examples, the initial experiments using traditional models achieved relatively high detection rates (81\%-85\%). Consequently, detection rates dropped to as low as 7\% (with a maximum accuracy of 53\% for RF) when adversarial examples were introduced. While the findings are relevant for understanding vulnerabilities in software-defined networks, the used dataset (CAIDA 2007 DDoS attack dataset) includes older traffic samples that might not resemble newer environments and techniques, limiting comparability with recent work featuring modern datasets. 
Abdelaty et al. ~\cite{Abdelaty-IEEE-2021} tackle this problem by evaluating their model on the CICIDS2017 dataset, a more comprehensive and diverse benchmark. Their study is also focused on generating high-quality samples and showcasing the impact of using generative models for generating network traffic against machine learning models used by tools for DoS/DDoS attack detection. The proposed architecture is based on two components. The first component is a generative model trained using normal traffic captures (WGAN-GP), aimed to generate additional traffic that respects the normal distribution of the features extracted from the training set. The role of the second component is to replace the features of normal traffic with features from the generation process. The authors show that including a GAN in the training stage of a model used to detect attacks leads to an F1-score improvement of 0.32, compared with the results obtained by the reference models.


\subsection{Adversarial training for DDoS detection}

Adversarial training is a neural network training technique that aims to enhance the robustness of neural networks by exposing them to adversarially perturbed data in the training phase. One way of achieving this is through dynamic adversarial training, which involves combining a primary objective function which needs to be minimized, with a secondary objective function (adversarial), which ought to be maximized. This training approach is meant to avoid learning certain examples or features that can affect the generalization of the model if they end up being learned by the model. Adversarial training requires careful application because it can affect the convergence of the model to an optimal point. For this reason, the optimization steps for minimizing the primary objective function are generally larger than the steps to maximize the secondary objective function. This can be achieved by weighting the objective functions so that the weight of the primary function is larger than that of the secondary function. Another way to achieve the same effect is by using different learning rates. Minimizing the primary objective function can be accomplished by applying the gradient descent algorithm. In theory, the secondary objective function is maximized by the gradient climbing/ascent. In practice, this behavior can be simulated by still applying the gradient descent algorithm, but changing either the sign of the objective function, the learning rate, or the gradients.

A practical example of dynamic adversarial training can be seen in the work of Zhang et al.~\cite{Zhang-ACMSIGSAC-2020}, who propose a deep learning-based defensive mechanism for IDS, called Tiki-Taka. In the first part of the paper, the authors test the vulnerabilities of three existing IDS models that use neural networks (MLP, CNN, and C-LSTM) against five recent black-box adversarial attacks. The result shows that up to 35.7\% of attacks successfully bypass the models, as evaluated on the CSE-CIC-IDS2018 dataset. The proposed improvement of the method targets three defensive mechanisms: model voting ensembling, ensemble adversarial training, and adversarial query detection. The first mechanism combines the performance of multiple classifiers to reduce the probability of an attack passing through the system unnoticed. The second one aims to continuously augment the dataset with adversarial examples and retraining the models. Finally, query detection adds another layer of security by blocking the attacker's IP address in case an attack is detected. The authors show that these methods are more efficient in their evaluations, with an increase in detection accuracy to almost 100\%.

In contrast, static adversarial training, as implemented by Nugraha et al.~\cite{Nugraha-IEEE-2021}, involves pre-generating adversarial examples (using mechanisms such as the FGSM and JSMA) before training and incorporating them into the dataset. The paper proposes a deep learning method to detect DDoS attacks, trained both with and without adversarial data. The authors employed two architectures, namely CNN-LSTM and MLP. When exposed to adversarial validation data, the MLP and CNN-LSTM models recorded drops of 11.99\% and 9.87\%, respectively. 
To address these performance drops, the authors attempted to train the MLP (chosen because of its efficiency) using three adversarial training procedures. Out of the tested methods, the one that involves replacing 80,000 examples from the training set with adversarial data achieves similar results to the originally obtained accuracy (99\%).


\subsection{Adversarial examples and attacks}

In the current literature, some methods for detecting and classifying DoS and DDoS attacks use different machine learning algorithms, including deep learning. With the exponential increase in traffic volume, algorithms such as SVMs, DTs, and Bayesian networks became inefficient for training and testing. Current directions in the architecture of network IDS favor deep learning methods because of their capacity to precisely classify traffic in a network, by learning certain abstract representations from a large volume of data.
However, since many of these models are trained on a single dataset, training sets and subsets originate from the same source. Thus, if the input data comes from an external source, with some small change in features, algorithms may fail to generalize. Since most machine learning algorithms are vulnerable to adversarial examples, lately, we have noticed attackers using generative AI techniques to generate them and produce incorrect classification decisions in standard detection systems. 
Even if the properties of an intrusion detection system are not known (black box), potential attackers can generate adversarial examples by repeatedly changing small subsets of features in the traffic (e.g., the time interval between consecutive packets). After each iteration, based on the received answer (confirmation/blocking of the attack or lack of an answer), the perturbations can be adjusted or changed completely until the network is breached. In this way, malicious fluxes are classified as benign traffic and remain undetected by the model.


The vulnerabilities of DoS IDS based on artificial neural networks against black-box adversarial attacks became an extensively studied matter. Peng et al.~\cite{Peng-IEEE-2019} propose an attack method that employs generating adversarial DoS data by disrupting relevant traffic features, achieving a synthetic data generation success rate of 80.77\%. Their experiments on KDDCup99 and CICIDS2017 datasets indicate high initial classification accuracies of 98.69\% and 93.45\%, respectively. These values drop significantly in the presence of adversarial examples. 
While Peng et al.~\cite{Peng-IEEE-2019} highlight the vulnerabilities of detection systems using synthetic adversarial data, Huang et al.~\cite{Huang-IEEE-2020} extend the approach by proposing two methods for generating DDoS adversarial data, which can both be directly applied to LSTM-based models: GA and PWPSA. The GA technique uses a genetic algorithm to gradually evolve derived data from the original data into DDoS adversarial attacks, with success rates between 72.50\% and 99.09\%. PWPSA employs an interactive process of modifying the data, determining the position and packet that affect the classification the most (based on a predefined function), with success rates between 67.17\% and 95.35\%. Overall, GA achieves better results and is also preferred in practice, while PWPSA is more computationally efficient, with a slightly lower success rate than GA.


One of the major challenges in training AI-based DDoS detection systems is the class imbalance typically present in network traffic datasets. In most real-world environments, benign traffic greatly outnumbers attack samples, which can bias machine learning models toward the majority class. To address this problem, several data augmentation techniques have been proposed. SMOTE \cite{Chawla-JAIR-2002} generates synthetic samples of the minority class by interpolating between existing attack samples in the feature space. This technique helps increase the representation of rare attack types without simply duplicating data points.
ADASYN \cite{He-IJCNN-2018} extends SMOTE by generating more synthetic samples in regions of the feature space where the minority class is sparsely represented. This allows the model to better learn hard examples, therefore leading to more robust decision boundaries between attack and normal traffic. More recently, GAN-based approaches have been explored for synthetic network traffic generation. Indeed, several recent studies have turned to GANs for both generating adversarial data and improving network IDS. Mustapha et al.~\cite{Mustapha-Elsevier-2023} showcases a method to detect DDoS attacks based on LSTM neural networks, which achieves 100\% accuracy on an initial dataset. This approach is further tested against adversarial attacks generated with a GAN model, where a significant decrease in performance is noticed. The final solution builds on the initial architecture by creating two new models: the first detects adversarial generated traffic, while the second classifies the traffic as normal or DDoS. The simultaneous use of the two techniques led to a final detection accuracy of 91.75\%. On a similar note, Shroff et al.~\cite{Shroff-Wiley-2022} initially use WGAN-GP (as Abdelaty et al.~\cite{Abdelaty-IEEE-2021}) to generate benign traffic captures, as well as DDoS attack captures, which are then used to test the accuracy of some classifiers. Furthermore, the data is used to experiment with a 5-layer deep network, achieving an accuracy of 95.37\% in identifying DDoS attacks generated by GANs. The authors show that manually modifying DDoS attack-specific features in benign data leads to an improved training method in the context of adversarial attack detection.

Recent work~\cite{Peng-IEEE-2019,Huang-IEEE-2020} has shown that machine learning models used in network intrusion detection are susceptible to adversarial attacks that manipulate input features in order to evade detection. In the context of DDoS detection, adversaries may craft malicious traffic flows whose statistical properties resemble legitimate traffic. This can be achieved by slightly perturbing packet-level or flow-level features, such as packet inter-arrival times, flow duration, or packet size distributions. Common adversarial attack techniques include gradient-based perturbation methods such as FGSM~\cite{Safaryan-ICML-2021} and JSMA~\cite{Wiyatno-ARXIV-2018}, which modify traffic features to mislead the classifier, while preserving functional traffic behavior. In network environments, adversarial examples may correspond to modified traffic patterns generated by botnets that deliberately mimic legitimate traffic bursts. The impact of such attacks on AI-based DDoS detection systems can be significant, as it may result in increased false negative rates, i.e.~malicious traffic that is classified as legitimate. To counter adversarial attacks, several model-hardening strategies can be employed. Adversarial training \cite{Abdelaty-IEEE-2021,Aleroud-CIW-2020,Nugraha-IEEE-2021}, where adversarial samples are injected into the training process, may improve robustness to adversarial attacks, but only to a limited extent. Other defense mechanisms involve ensemble learning \cite{Zhang-ACMSIGSAC-2020}, where multiple detection models can be integrated to reduce susceptibility to single-model attacks, or dimensionality reduction \cite{Wei_ACCESS_2021}, where the sensitivity of classifiers to small perturbations is reduced by keeping high-variance features. Such approaches improve the robustness of AI-based DDoS detection systems and represent an important direction for future research.


He et al.~\cite{He-IEEE-2023} find that feature-level attacks such as FGSM and JSMA do not generate practical adversarial data. Furthermore, dependence on outdated datasets, like SL-KDD and KDDCup99, decreases the relevance of results. More recent datasets, such as CICIDS2017/2018 provide better benchmarks, but indicate the need for further development in adversarial traffic generation. White-box attacks are highly effective in evading detection systems, but black-box attacks still struggle because of the increased traffic complexity. As also noted by Abdelaty et al.~\cite{Abdelaty-IEEE-2021} and Saka et al.~\cite{Saka-TC-2023}, GAN-based architectures show potential in adversarial data generation. Nevertheless, He et al.~\cite{He-IEEE-2023} conclude that the literature still needs more representative and diverse datasets, along with more robust defense mechanisms such as adversarial training and feature reduction strategies.

\section{AI Generated Mitigations}\label{sec:ai-mitigation}

Once a DDoS attack is mounted,
we are continuously pressured to act and to act fast.
After detecting the attack,
we quickly have to mitigate it through various techniques, most of which involve blocking the attack nodes while allowing legitimate traffic to navigate as usual.
The complexity of existing attacks and network topologies
requires us to write sophisticated and efficient blocking rules:
we want to block as many connections as possible within a single firewall rule
such that throughput is not throttled.
The last couple of years have shown an emergent trend
that passes the responsibility of compiling block lists and composing efficient filtering rules
from cybersecurity agents to specialized AI-agents that build custom-made DTs
or fine-tune LLMs to generate firewall rules.

\hlight{R1-C6}{
Table \ref{tbl:mitigation-comparison} summarizes the main operational trade-offs across the AI-based mitigation families covered in this survey. Since the literature is still relatively recent and heterogeneous, the table is intentionally qualitative and is meant to complement, rather than replace, the detailed discussion that follows.

\begin{table}[t]
    \centering
    \begin{tabular}{p{1.5cm}p{2.5cm}p{1.3cm}p{1.0cm}p{1.3cm}p{3.5cm}}
        \toprule
        Family & Outputs/actions & Comp. Effort & Latency req. & Scalability & Typical failure points \\
        \midrule
        DT \cite{Zadnik-IIS-2023, Coscia-JISAS-2024} & Interpretable filtering rules (e.g., Suricata/iptables-style rules) & Low & High & High & Overspecialized rules, collateral damage when thresholds or features are poorly chosen \\
        \addlinespace \midrule
        RL \cite{Simpson-TNSM-2020, FENG-IWQOS-2020, Li-TITS-2022, Kumar-JIT-2024, JAYAKRISHNA-RIE-2025, DUAN-SEAMS-2025} & Rate limit, rerouting, redirection, challenge/puzzle selection & Medium at inference, High at training & High & Medium & Reward misspecification, unstable training, discrete and scenario-specific action spaces, large training budgets \\
        \addlinespace \midrule
        LLM \cite{Louro-ARS-2024, Yin-ARXIV-2024, Wang-APNet-2024} & Firewall/IDS rules and templated mitigation commands & High & Medium to High & High & Hallucinated rules, no syntactic/semantic validation, large memory and accelerator requirements \\
        \addlinespace \midrule
        Human-in-the-loop \cite{Wang-APNet-2024, Paduraru-ICSOFT-2024} & Candidate rules, explanations, and operator guidance & Medium & Medium & Medium & Analyst bottlenecks, inconsistent review speed, dependence on expert availability \\
        \bottomrule
    \end{tabular}
    \caption{Qualitative comparison of the main AI-based DDoS mitigation families discussed in this survey. The labels Low/Medium/High are relative to the surveyed literature and summarize training/deployment computational effort, latency requirements, scalability, and we also describe the typical operational failure modes.}
    \label{tbl:mitigation-comparison}
\end{table}}

Initially, several AI-based DDoS mitigation techniques were based on Reinforcement Learning (RL), or Deep Reinforcement Learning (DRL). The setup is natural: the state represents current and past information about the network packet/flow traffic, the reward function uses the DDoS detection models to establish the current level of attack (network throughput, for example), and the actions are a set of pre-defined network parameters that can be controlled in order to reduce the DDoS attack effectiveness while minimizing collateral damage and control costs. Among the first research directions, the work in \cite{Simpson-TNSM-2020} uses RL as an online feedback controller for per-host and per-flow traffic throttling. The RL agent combines global traffic load measurements and per-flow statistics to decide the probability $p$ of dropping packets for a particular host/flow pair. The reward function encourages keeping the received legitimate traffic volume high, and other quality indicators of the network. The authors do point out the difficulties of choosing appropriate reward functions and suggest several future research directions based on this observation. Soon after, the work in \cite{FENG-IWQOS-2020} used RL to directly choose mitigation actions per incoming application-layer requests. The RL mechanism decides among a fixed number of actions, which describe ways the requests are processed (partially, delayed, or fully) or are blocked. The reward function uses several performance indicators to decide system occupation rate (CPU, memory, and link utilization), and these are used to decide future actions. A Q-learning style value iteration, with a deep neural network, uses the occupancy rate features to decide among the discrete set of actions.

These RL mitigation techniques were then extended to mobile and IoT networks. A method called FAST \cite{Li-TITS-2022} was used to mitigate DDoS attacks on Multi-access Edge Computing (MEC) base stations. The model uses a Kalman filter approach to estimate near-future traffic and evaluate the severity of the attacks numerically. In the RL mechanism, this quantity is used as a reward function that indicates the scale of the DDoS attack. This is coupled with Q-learning, assisted by an asynchronous DDQN that provides better early action guidance; later, the method falls back to $\epsilon$-greedy Q-learning exploration. Actions are taken again from a discrete, pre-defined set. For IoT networks, Message Driven Reinforcement Learning (MD-RL) \cite{Kumar-JIT-2024} was used to mitigate DDoS attacks by avoiding and isolating malicious nodes of the network, by using Ad hoc On-Demand Distance Vector (AODV). In this context, mitigation does not mean that packets/flows are dropped; rather, the model forwards the data and retransmits it via alternative routes. This is achieved by continuously monitoring packet forwarding in the IoT network and deciding which next hop a packet should traverse. The reward function is again driven by classic network performance indicators such as throughput, latency, etc.

The latest research efforts in this DRL-based direction focus on establishing real-time mitigation and utilizing more sophisticated learning models. Reinforcement Learning-based Adaptive Rate Limiting (RL-ARL) \cite{JAYAKRISHNA-RIE-2025} dynamically adjusts in real-time the traffic rate limits based on the severity of the detected DDoS attacks, effectively mitigating the impact of the attack. Rewards/penalties are such that they encourage blocking malicious requests while preserving performance, i.e., lower latency, higher throughput, fewer disruptions to benign traffic, etc. And finally, the work in \cite{DUAN-SEAMS-2025} introduces the DosSink model, which integrates detection and mitigation through Variational AutoEncoders (VAE) and actor-critic DRL. The detection mechanism gives each traffic flow a risk score, and then the DRL agent chooses the mitigation techniques among a list of possibilities (traffic limiting and redirection, or puzzle-based source verification actions to slow down the attackers). True positive rates and false positive rates are separately explicitly stated and are excellent, but the number of training episodes is sometimes large, exceeding 20.000.

Recent work tries to overcome the limitation of DRL-based methods, where the mitigation action set is usually discrete and limited in size, by using modern LLMs. This approach greatly generalizes the use of automatic AI-based mitigation techniques, but comes with its own drawbacks. Louro et al.~\cite{Louro-ARS-2024}
establish a baseline by employing existing LLMs,
both commercial and open-source,
to tackle the mitigation task,
a task at which they show limited to no success rates.
Here,
mitigation consists of emitting proper \texttt{iptables} and \texttt{snort} rules for various DDoS attacks that are passed as (structured) prompts.
Significant improvements are shown by employing
LoRA of LLMs
and
PEFT
techniques
to fine-tune existing open-source models, Mistral and LLaMA,
for this mitigation task,
which result in outputs consisting of advanced filtering rules
with high success rates.
Following the same process,
ShieldGPT~\cite{Wang-APNet-2024}
provides an AI-based DDoS mitigation software architecture
that classifies incoming DDoS attacks
and fine-tunes GPT for prompt templating
\texttt{iptables} rules as required by each attack type.

Some approaches,
such as DrLLM~\cite{Yin-ARXIV-2024},
avoid fine-tuning
by employing prompt engineering techniques like
CoD,
for templating the output,
and
Zero-shot CoT,
for processing incoming flows as online time series.
The authors prove the success of this approach
on popular vanilla models
like GPT, LLaMA, Qwen2 and Deepseek.
While DrLLM
handles solely DDoS detection,
regardless of the attack type,
this approach could be further extended
to the mitigation tasks discussed above.

If so far we have seen flows based on
tools for tools,
LLMs that produce rules for packet filters,
there is also literature 
on fine-tuning LLMs for human agents under DDoS attack.
Indeed,
in ShieldGPT~\cite{Wang-APNet-2024},
the authors design a
a separate sub-task
to provide an explanation prompt template
for the cybersecurity agent handling the case.
The work of Păduraru et al.~\cite{Paduraru-ICSOFT-2024}
introduces CyberGuardian based on a fine-tuned LLaMA model
in order to
aid agents in tackling various cybersecurity attack scenarios.
In the DDoS scenario,
23 agents participate in simulated scenarios
where they prompt the LLM
to achieve two tasks:
generating IP block lists
and
emitting appropriate firewall commands to stop the attack.

A very popular ML technique to learn AI generated mitigations to DDoS attacks is DTs. These models are very popular in our setting because they are highly interpretable providing transparent verdicts, their results can relatively easily be transformed into logical rules based on \textit{and}/\textit{or} operations, and they can be trained very fast as compared with other ML techniques from the literature.
The work in~\cite{Zadnik-IIS-2023} proposes a DT model to infer filtering rules to mitigate volumetric DDoS attacks. The authors explain how to convert the DT model into interpretable logical rules and they obtain state-of-the-art results. The work in~\cite{Coscia-JISAS-2024} proposes a method called Anomaly2Sign which is based on DT models and generates Suricata rules for a wide range of DDoS attacks. The authors also focus on training the simplest DT models, measured by AIC scores, in order to further improve the interpretability and simplicity of the learned rules. To train their DT models, both papers assume that the available dataset is labeled and separated (albeit not exactly perfectly) into legitimate and illegitimate traffic data. In addition, both papers highlight the low training times, below one second, for their DT models and tout their computational efficiency compared with other ML models. This is a crucial factor when considering real-time or online training scenarios. Finally, both papers report near-perfect mitigation results for their proposed methodologies.

The problem of AI generated mitigations to DDoS attacks has only recently been approached by the research community, and therefore, the state-of-the-art literature is thin. Further developments are expected in the years to come as a response to the fact that DDoS attacks are themselves enhanced by AI techniques. Still, we already observe that some characteristics specific to AI generated DDoS mitigation are starting to take shape.

First, new mitigation quality indicators have been introduced to evaluate the effectiveness of the proposed solutions. In time-sensitive scenarios, one of the most important indicators is the Time To Mitigation (TTM), which is the elapsed time from AI model detection to enforcing a new mitigation rule. Anomaly2Sign reports TTMs below 1.5s and \cite{Zadnik-IIS-2023} reports below 7s, while other papers do not report TTMs. Then, similar to the Recall indicator, the Residual Attack Traffic (RAT), i.e., the fraction of packets/flows that still reach the victim network after the rule enforcement was completed, is used and is independent of the quality of the AI detection model. Equally important, similar to the False Positive Rate, the Collateral Damage (CD) factor measures the fraction of benign traffic that is blocked by the newly enforced rules. Together, RAT and CD establish the quality and the effect of the rules enforced in the mitigation pipeline. Finally, the Rule Complexity (RC) indicator evaluates the quality of the mitigation rules emitted: the number of rules, their complexity (length or number of parameters involved), etc. The goal is to keep RC low in order to keep generalization, explainability, and latencies under control. Larger, brittle rule sets that are not able to successfully generalize to new, similar but not identical traffic situations need to be identified and avoided. In this vein, Anomaly3Sign makes the most valiant efforts to minimize the rule count by formalizing and enforcing the concept of an optimal rule set.

To measure and validate these new indicators in experimental settings, recently published papers create DDoS simulation scenarios where labeled traffic is used to create complex mixed attack scenarios, where RAT and CD can be computed. Both Anomaly2Sign and ShieldGPT define sophisticated mixed DDoS attacks using labeled data, while \cite{Zadnik-IIS-2023}, under the same testing circumstances, even considers wrong labeling in order to test the robustness of the mitigation pipeline. To establish low RC and to guarantee that the mitigation rules are safe, ShieldGPT checks the automatically generated rules syntactically and semantically, whereas \cite{Louro-ARS-2024} has the rules checked by an additional expert human operator before enforcement. In the latter case, the long-term goal may be to use RL to train future automatic rule checkers from human feedback and corrections.

Finally, we highlight that all AI mitigation techniques which are discussed also entail their own drawbacks and risks. Following the ``first, do no harm'' mantra, in most cases, the biggest practical concern is that of self-inflicted outages caused by harsh or incorrect mitigation rules. This situation would defeat the whole purpose of having an automated DDoS detection and mitigation solution. Second, an important risk is that of increased processing effort and latencies in the overall detection pipeline, which would negatively affect the accuracy and timeliness of DDoS attack detection. As a last point of concern, we have to mention that any mitigation solution currently based on LLMs, such as \cite{Louro-ARS-2024, Yin-ARXIV-2024, Wang-APNet-2024}, may be subject to incorrect decisions due to model hallucinations, and therefore, extra processing steps to validate the mitigation rules should be planned. All previously defined indicators, TTM, RAT, CD, and RC, should be used to decide if the mitigation system is anywhere near these risk scenarios and take appropriate actions, including human-in-the-loop decisions.

\hlight{R1-C7}{
From a deployment perspective, these mitigation techniques span a wide computational spectrum. Decision Trees approaches are currently the lightest and appear closest to practical deployment, since both training and inference are fast \cite{Zadnik-IIS-2023},\cite{Coscia-JISAS-2024} and the generated rules are interpretable \cite{Coscia-JISAS-2024}. Reinforcement Learning methods typically have computationally complex training phases \cite{Simpson-TNSM-2020} and are therefore attractive when repeated offline simulation is possible. In general, these methods still require careful engineering to keep online action selection and state updates within tight latency budgets \cite{FENG-IWQOS-2020}. LLM-based approaches are the most demanding in terms of memory footprint and rule-validation overhead, which currently makes them more suitable for offline analysis assistance rather than for online deployment. For these reasons, operational constraints should be considered jointly with TTM, RAT, CD, and RC when evaluating whether an AI-based mitigation pipeline is deployable in production.
}

\section{Ethical Considerations and Societal Impact}\label{sec:ethics}

We briefly discuss several ethical and societal considerations when dealing with cybersecurity and AI in general, and in particular, how it reflects on their use in
detection and mitigation of DDoS attacks.

\noindent\textbf{Dual-use nature of adversarial techniques:}
Although we discuss adversarial approaches
in the context of training and improving the robustness of learning based DDoS detection models in Section~\ref{sec:ai-traffic},
the generated examples and attacks can also be used for employing real attacks.
This dual-use nature of adversarial techniques is a well known ethical dilemma and its use in defense scenarios, such as the ones described in this survey,
has to be complemented by well-defined policies and regulations that confine its use to the greater good.

\noindent\textbf{Fairness risks and discriminatory false positives:}
AI-based models heavily rely on the trained dataset, while inferring data from a potentially distinct distribution in production.
This leads to both false positives and false negatives in practice. While false negatives allow DDoS traffic to pass through,
false positives block legitimate users, thus discriminating against fair use.
In order to minimize this risk,
production traffic should be used to retrain and specialize the model
on network data patterns that are specific to the institution being protected against these types of attacks.
We note that the task of eliminating false positives
is especially hard in the DDoS context.
First, we deal with a greatly unbalanced setting,
i.e.~there are many more normal samples than attack samples.
Second, the attacks generally follow a Laplace distribution model,
where the peaks represent dense targeted attacks that, for example,
can take from a few hours to a few days in a year,
with the rest of the time encountering strictly legitimate traffic.

\noindent\textbf{Privacy implications of traffic analysis:}
Traffic analysis,
especially real traffic that is recorded or analyzed in real-time,
can have privacy implications depending on how the data is preprocessed and stored.
To avoid misuse of user information and behavior,
such as targeted monitoring and disclosure of private data,
DDoS detection and mitigation tasks should employ
dedicated data preprocessing policies to enforce this.
For example,
hardware or software TAP devices are commonly used for traffic recording.
These can be configured to retain just the packet headers (without the payload), thus avoiding the storage of actual user data.
Moreover,
for TCP streams, we can have the packet headers aggregated into connection statistical data, called flows in the literature,
thus further obfuscating the actual user meta-data.
Regardless of the preprocessing techniques,
data collection practices should be made transparent to the end users,
by the institution employing the detection and mitigation solutions.

\noindent\textbf{Environmental costs of large-scale AI models:}
Large-scale AI models have been used for the mitigation task (see Section~\ref{sec:ai-mitigation}).
It is well known that these models require large amounts of data
and employ significant amounts of computation during training,
leading to high energy consumption with significant environmental costs due to carbon emissions.
Nonetheless,
in the surveyed literature, we have not identified existing foundational models
or other large-scale AI models that have been specifically trained for the mitigation task.
Current work resumes to using existing models (just inference, templated or not),
while few take the extra step of fine-tuning these models through standard methods (e.g.~LoRA).
This has a limited environmental cost.

\noindent\textbf{Safety concerns related to AI-generated mitigation rules:}
Care must be taken when relying on AI-generated mitigation rules,
as most are based on LLMs that are known to hallucinate.
Indeed,
this can lead to service disruptions and outages through blocking legitimate ingress and egress traffic.
These solutions need to be assisted by continuous monitoring and error-correction protocols involving expert network operators.

As described above,
tackling these
ethical considerations along with their potential societal impact
is an essential part of the emerging dominance of AI-based cybersecurity solutions in general and of the anti-DDoS field in particular.

\section{Conclusions and Future Research}
\label{sec:conclusion}

In this survey,
we provide a comprehensive review
of the most popular AI-based detection and mitigation techniques
for volumetric, protocol, application,
reflection and amplification attacks.
While DDoS attack and mitigation techniques vary wildly across time, victim services, network infrastructure, computer systems and attack delivery, the study introduces a clear definition of DDoS attacks
and offers a manual and automatic taxonomy
of existing research.

Available datasets are discussed together with 
alternative data formats 
such as
time series, graphs, and tabular data formats
which are meant to aid the learning process. In fact, we find that in most cases, the data format has a large impact on the learning algorithms used. Published detection rate results suggest that most currently available datasets lack the complexity of real-world situations and therefore, even classic machine learning methods saturate them.

Of special note are the sections that treat
AI generated attack mitigations
and
AI generated traffic for adversarial training, examples and attacks. The survey finds that AI-driven mitigation remains less mature than detection, but the most effective strands are converging on rule-producing approaches that can be operationalized. Most notable, interpretable models that translate naturally into firewall rules, alongside early work leveraging LLMs to assist human operators and automate aspects of rule generation. 

This work
also leads to multiple insights into future research venues that we describe below.

\noindent\textbf{Cross-dataset testing:}
A remaining open question concerns how well is the detection quality maintained across new, potentially unseen datasets. 
The risk of literature overfitting is valid, given the relatively small number of popular datasets used in most papers,
and although private datasets are sometimes reported in the literature, 
the sensitive nature of network traffic and the difficulty of capturing relevant attacks
are significant impediments.
A recent investigation \cite{Anley-CS-2024} reports 
drops in detection accuracy in the range of 5\% - 10\% for different models,
when evaluating on datasets other than the ones used for training. 
Solutions based on supervised learning are, admittedly, affected to a greater extent compared to unsupervised approaches.
Yet, the issue is largely under-addressed in the literature, 
especially considering the impact it may have when deploying anti-DDoS systems in the wild.

\noindent\textbf{Anti-DDoS tailored algorithms:}
Our literature review has shown limited specialization or data adaptation regarding AI algorithms.
Most of the work was focused on handling cybersecurity attacks in bulk, as they were found in the various available datasets
and the few works that specialized in DDoS attacks
used standard shallow or deep learning algorithms
for the task.
We consider that future directions can improve results in terms of accuracy and, more importantly, detection and mitigation times,
if new anti-DDoS tailored algorithms are investigated.
Improvements will probably be more visible when coupled with cross-dataset testing.

\noindent\textbf{Hybrid models for improved detection:}
In order to minimize the False Positive Rate of the detection systems, different hybrid models can be employed for each category of DDoS attacks. In case of volumetric attacks, which are characterized by a sudden increase in network traffic volume, simple statistical methods like EWMA or rolling window robust Z-Score can be combined with anomaly detection models based on IF or LSTM autoencoders. The detector corresponding to each type of attack would require a different set of relevant features for optimal performance. For example, the detection of SYN floods can leverage the rolling window robust Z-Score to detect spikes in the time series determined by the number of SYN flags encountered in fixed-length time windows. The final score can be derived using either hard or soft scoring from the individual detectors.

Multi-modal systems can be designed to exploit information from both layer 3 and layer 7 when detecting application DDoS attacks. We assume that logs from web servers and resource usage metrics can augment the information extracted from the network level, enabling a more accurate detection.

\noindent\textbf{Dynamic data format adaptation:}
DDoS attack bandwidths can vary widely.
While most are centered around 1Gbps,
recently we often find 5-10Gbps reports in the wild.
A common attack behavior is that DDoS start small and grow toward their peak,
as more bots are enabled, reflection and amplification are enabled, etc.
Current literature does not cover how AI algorithms
should handle this ramp-up and how traffic data storage and processing impacts detection and mitigation quality.
For instance, one can easily imagine that storing raw packet data or even flows when under 10Gbps DDoS attacks
is an almost impossible task:
storing each packet (even in memory) for AI inference purposes can lead to self-DoS-ing
the in-memory or on-disk database.
Instead,
algorithms need to be able to adapt
and handle multiple data formats depending on the attack size:
for example we can keep raw data until the attack reaches 1Gbps,
switch to flows afterward until 2Gbps,
and then shrink the data from IP-level to IP-class level
until 5Gbps,
and finally switch to a handful of statistics that can be quickly computed, stored and inferred.

\noindent\textbf{Detection and mitigation time for algorithm quality assessment:}
Catching DDoS attacks early on is critical to keeping the victim infrastructure running.
Still,
almost the entire literature tackles this task as a standard classification problem:
choose a dataset,
train a classifier,
test and validate the results
and report standard metrics such as
accuracy, F1-score, area under the curve, true positive rate,
true negative rate, etc.
While this is indeed necessary for basic model validation,
we consider that the domain-specific validation should be
detection time coupled with mitigation time and effectiveness, where applicable.
More on mitigation,
we consider that besides mitigation time,
metrics regarding normal infrastructure operation
and regular traffic throughput while under DDoS
should be measured and reported by future research
when proposing new mitigation algorithms and methodologies.

\noindent\textbf{Fuzzy detection:} Most solutions use AI to partition the traffic into two distinct classes: normal and attack.
In contrast,
the prediction of ML algorithms is in fact a probability score,
and the binary class label is determined via a probability threshold.
Based on this observation,
the detection threshold can be dynamically adjusted in order to maximize service availability.
For example,
when a server load is low,
one can decrease the threshold
such that more traffic is passed through and false positives are reduced.
Subsequently,
during high load the threshold can be increased.
From another perspective,
this can be seen as a scheduling algorithm
where the classification score acts as traffic priority.

\noindent\textbf{Comprehensive DDoS dataset:}
Most public datasets contain network recorded DDoS traffic
covering a small subset of attack types, often mixed with non-DDoS attacks.
We consider that there is a need for dedicated datasets
that encompass all known DDoS attack types as well as more unusual benign traffic, like games, various messaging services, peer-to-peer exchanges, blockchain and SSH, to name a few.
The traffic should be labeled and 
include further annotations regarding current throughput,
bots details, 
reflection and amplification.
The datasets should include various bandwidth traffic shapes, such as
ramping up (e.g.\ from 100Mbps to 10Gbps),
maintaining a fixed bandwidth (e.g.\ 5Gbps for 30 minutes),
sine behavior (e.g.\ up and down from 1Gbps to 3Gbps for 2 hours).
Expert mitigation rules that can be employed at various times to handle the attack should also be provided in order to evaluate mitigation algorithms and the firewall rules they provide.

Furthermore, we identify the need to have more complex and realistic datasets that mimic the real-world practical difficulties of detecting DDoS attacks. According to classic AI performance metrics, many current AI-based detection methods perform nearly perfectly on the available datasets. We believe that, at least partially, these results also betray the simplistic nature of the available datasets. Finally, training detection methods on current datasets do not generalize well to new DDoS attacks, even when these are variations on classic DDoS attacks. It is therefore desirable to have an adversarial approach to dataset design such that we do not overfit detection methods for particular scenarios and we thus prepare for real-world scenarios.

\noindent\textbf{Adversarial training:} Adversarial training is another area that is insufficiently explored. We consider that adversarial training techniques could broaden the applicability and robustness of AI-based DDoS attack detection methods to out-of-distribution samples. For example, in computer vision, adversarial domain adaptation \cite{Tzeng-CVPR-2017} was shown to boost results in cross-domain settings. We consider that such an approach is also likely to perform well in DDoS attack detection for cross-dataset scenarios.

\noindent\textbf{Robustness against adversarial traffic:} Future research should focus on improving the robustness of AI-based DDoS detection systems against adversarial traffic manipulation. Current evaluation protocols often rely on static datasets and do not consider adaptive attackers that modify traffic patterns to evade detection. Important directions include: developing benchmark datasets containing adversarially-perturbed traffic, designing adaptive detection models capable of online learning, and evaluating robustness under realistic traffic distribution shifts. Improving robustness is essential for deploying AI-based detection systems in operational networks where attackers continuously adapt their strategies.

\noindent\textbf{Explainable AI:} An understudied topic in the area of DDoS attack detection is the development of explainable AI methods. As in other domains, the success of deep learning methods comes with an important downside, namely that the reasons behind the predictions are often unknown or hard to precisely determine. The literature in this direction is scarce \cite{Alzubi-CMC-2024,Das-IEMCON-2021}, lacking sufficient exploration towards explainable DDoS attack detection based on various types of deep neural networks. Knowing the reasons behind a decision could also be useful in the mitigation phase.

\noindent\textbf{AI mitigation:}
AI generated firewall rules are still an early research topic.
While existing work has shown that it is a valid pursuit,
much work can be done in this direction.
Currently, algorithm efficiency is estimated by comparing the generated outputs with expert written rules for specific one-time attacks.
We consider that this can be further improved by
continuously generating and updating firewall rules,
while maintaining the context of a prolonged attack.
During the attack, the algorithm has to take into consideration
updating, merging or removing existing rules,
not just inserting new ones
(e.g.\ we want to block an entire IP class instead of keeping 255 separate block rules).
The main goal here is to keep the firewall chain rules efficient, such that throughput is optimized;
the victim infrastructure has to be kept running and regular traffic has to pass through as close to the network's nominal functionality as possible.

We recognize that a few recent studies have explored the use of LLMs to automatically generate mitigation rules once a DDoS attack has been detected \cite{JAYAKRISHNA-RIE-2025,Yin-ARXIV-2024}. The effectiveness of such approaches strongly depends on the prompt design used to guide the model. One promising strategy is Chain-of-Thought (CoT) \cite{Yu-ARXIV-2023} prompting, where the model is encouraged to reason step-by-step before producing a firewall rule. For example, the prompt may instruct the model to identify suspicious traffic characteristics, determine the attack type, and derive a corresponding filtering rule. This reasoning process can improve the consistency and correctness of generated mitigation rules. Another strategy is Constraint-of-Deviation (CoD) \cite{Liu-CHI-2024} prompting, which explicitly constrains the generated rules to satisfy predefined network policies. For instance, a prompt may require that the generated rule only blocks traffic exceeding a certain packet rate or originating from suspicious IP ranges. More importantly, it may include constraints that make the rule valid. Prompt engineering techniques can therefore play a critical role in ensuring that LLM-generated mitigation strategies are both effective and compliant with network security policies.

\backmatter


\section*{Declarations}

\bmhead{Funding}
The authors were funded by a grant of the Ministry of Research, Innovation and Digitization, CCCDI - UEFISCDI, project number PN-IV-P6-6.3-SOL-2024-2-0197, within PNCDI IV.
This research is also supported by the project ``Romanian Hub for Artificial Intelligence - HRIA'', Smart Growth, Digitization and Financial Instruments Program, 2021-2027, MySMIS no.~351416.

\bmhead{Conflict of Interest}
The authors declare no Conflict of interest.

\bmhead{Ethics approval and consent to participate}
Not applicable.

\bmhead{Consent for publication}
Not applicable.
\bmhead{Data availability}
Not applicable.
\bmhead{Materials availability}
Not applicable.
\bmhead{Code availability}
Not applicable.

\bmhead{Author contribution}
P.I. wrote and oversaw the main manuscript text, wrote most of Sections 1, 2, and 8 and prepared Tables 1, 2, 5, 6 and Figures 3, 6.
R.I. wrote Section 3.3, supervised Section 6 and 7 and generated Figure 4.
C.R. supervised Sections 4 and 5 and prepared Figures 1, 2, 5 and Table 3 and 7.
A.A. researched and wrote the initial draft of Sections 6 and 7 and prepared Figure 7.
S.G. researched and wrote the initial draft of Section 5.4 and prepared Figures
1, 2 and Table 3.
A.H. researched and wrote the initial draft of Sections 4, 5.1, 5.3 and prepared Figure 2.
A.P. researched and wrote the initial draft of Section 4.
N.C. researched and wrote the initial draft of Section 5.2 and prepared Figure 3.
All authors reviewed the manuscript.

\bibliography{bib}
\end{document}